\documentclass[11pt,a4paper]{article}
\usepackage{jcappubnew}

\newcommand{\beq}{\begin{equation}}
\newcommand{\eeq}{\end{equation}}
\newcommand{\bea}{\begin{eqnarray}}
\newcommand{\eea}{\end{eqnarray}}

\newcommand{\vc}[1]{{\boldsymbol #1}}

\newcommand{\lh}{\left(}
\newcommand{\rh}{\right)}
\newcommand{\der}{\partial}
\renewcommand{\d}{\mathrm{d}}
\newcommand{\non}{\nonumber}
\DeclareMathSymbol{\mg}{\mathrel}{symbols}{"1D}

\newcommand{\gd}{\delta}
\renewcommand{\ge}{\epsilon}

\newcommand{\get}{\eta}

\newcommand{\gk}{\kappa}

\newcommand{\gx}{\xi}

\newcommand{\gf}{\phi}

\newcommand{\gc}{\chi}

\newcommand{\bv}{{\bar v}}

\newcommand{\bH}{{\bar H}}

\newcommand{\bN}{{\bar N}}

\newcommand{\ha}{{\hat \alpha}}
\newcommand{\hr}{{\hat \rho}}

\newcommand{\bdm}{\begin{displaymath}}
\newcommand{\edm}{\end{displaymath}}
\newcommand{\nn}{\nonumber}
\def\be{\begin{equation}}
\def\ee{\end{equation}}

\def\e{\epsilon}
\def\es{\epsilon_*}
\def\hpa{\eta^{\parallel}}
\def\hpe{\eta^{\perp}}
\def\hpas{\eta^{\parallel}_*}
\def\hpes{\eta^{\perp}_*}

\def\ae{\alpha_{(1)}}
\def\aee{\alpha_{(2)}}
\def\tvarphi{\tilde{\varphi}}
\def\tdvphi{\dot{\tilde{\varphi}}}

\def\dvphi{\dot{\varphi}}
\def\dphi{\dot{\phi}}

\title{Gauge-invariant perturbations at second order in two-field inflation}

\author{Eleftheria Tzavara}
\author{and Bartjan van Tent}
\affiliation{Laboratoire de Physique Th\'eorique, Universit\'e Paris-Sud 11 
and CNRS,\\ B\^atiment 210, 91405 Orsay Cedex, France}
\emailAdd{Eleftheria.Tzavara@th.u-psud.fr} 
\emailAdd{Bartjan.Van-Tent@th.u-psud.fr}

\abstract{We study the second-order gauge-invariant adiabatic and isocurvature perturbations in terms of the scalar fields 
present during inflation, 
along with the related fully non-linear space gradient of these quantities. 
We discuss the relation with other perturbation quantities defined in the literature. 
We also construct the 
exact cubic action of the second-order perturbations (beyond any slow-roll or super-horizon approximations and including 
tensor perturbations), both in the uniform energy-density gauge and the flat gauge in order to settle various 
gauge-related issues. 
We thus provide the tool to calculate the exact non-Gaussianity beyond slow-roll and at any scale.}

\begin{document}

\begin{flushright}
LPT-11-102 \\ 
\end{flushright}

\maketitle
\flushbottom

\section{Introduction}

The concept of inflationary curvature perturbations was first invoked in order to explain the primordial fluctuations that 
source the CMB anisotropy and structure formation 
\cite{paper1,paper2,paper3,paper4}. The inflationary paradigm has been observationally tested for more than 
10 years and its prediction for an almost scale invariant 
spectrum of the first-order curvature perturbations has been verified by the data of many experiments (see for example 
\cite{Komatsu:2010fb}).
Since the definition of perturbations depends in general on the gauge choice, a gauge-invariant definition of the 
cosmological perturbations 
is of vital importance to make contact with physical observables, which are obviously gauge-invariant. 
That was investigated in detail in \cite{Bardeen:1980kt} and later in \cite{Mukhanov:1990me}. 
In the mean time the need for more observational quantities than just those based on linear perturbation theory  
has become clear, in order to break the degeneracy of the immense number of 
inflationary models. One of the most fruitful has proven to be the non-Gaussian 
characteristics of the perturbations. This has led to the development of new methods 
to study the combination of the scalar field and metric 
perturbations, the one sourcing the other, in a gauge-invariant way beyond first order.

It was not until 2003 that Malik and Wands in \cite{Malik:2003mv} defined the gauge-invariant quantity at second order that reduces to the 
curvature perturbation in the uniform energy-density gauge. In \cite{Malik:2005cy} the super-horizon equations of motion of 
these quantities were derived (but see also \cite{Noh:2003yg} for a gauge-ready formulation of the 
perturbations and their equations). Another way to deal with perturbations at second order is the space gradients approach 
first introduced in \cite{Ellis:1989jt} and specifically the gradient of the fully non-linear curvature perturbation used in 
\cite{Rigopoulos:2005xx} and defined by the same authors in \cite{Rigopoulos:2004gr}. The advantage of the method is that when 
the space gradients are expanded to first order they are automatically gauge-invariant. Some years later the gradient of the  
curvature perturbation was redefined in a covariant way in \cite{Langlois:2005qp}. In this paper it was shown that 
when expanded to second order, this quantity reduces to the gauge-invariant curvature perturbation defined in \cite{Malik:2003mv} 
plus a gauge transformation term. 

In this paper we generalize the definition of the gauge-invariant curvature perturbation (or the gradient 
of the relevant fully non-linear quantity) in terms of the energy density to a definition in terms of the scalar fields present 
during inflation and study the consequences of this change at second order. 
Since the scalar fields are the principal 
quantities during inflation, it makes more sense to use these as a starting point, especially in the case of multiple-field 
inflation.
Our original need for such a definition was to find the 
horizon crossing contribution to the second-order curvature perturbation in terms of the first-order ones 
in the long-wavelength formalism \cite{Tzavara:2010ge}.
Indeed, such a definition helps to fully understand and potentially generalize the two formalisms used to compute non-Gaussianity 
during inflation, 
i.e. the long-wavelength formalism \cite{Rigopoulos:2005xx,Rigopoulos:2011eq,Tzavara:2010ge} and the $\delta N$ formalism 
\cite{Starobinsky:1986fxa,Sasaki:1995aw,Sasaki:1998ug,Lyth:2004gb,Lyth:2005fi}, 
where instead of the energy density, the values of the fields themselves are used. 

In the case of multiple-field inflation not only an adiabatic curvature perturbation is produced, but also one or more 
isocurvature perturbations. 
We would like to provide the same type 
of study for the isocurvature perturbation as for the curvature one, using the generalized quantity defined in \cite{Rigopoulos:2005xx}, and 
deduce from that 
the second-order gauge-invariant analogue. 
This definition of the isocurvature perturbation makes direct contact with the scalar fields during 
inflation (instead of using their pressure), which we find more useful during the period of scalar field domination of the universe. 
It has a simple physical meaning, 
that is the combination of the fields that remains orthogonal to the field trajectory, as opposed to the adiabatic perturbation that is 
parallel to the field trajectory (and proportional to the energy density).

On a related subject, Maldacena in \cite{Maldacena:2002vr} found the third-order action for the first-order 
adiabatic perturbation in a single-field dominated universe, both in the flat gauge and in the uniform energy-density gauge. 
In order to rewrite the action in a gauge-invariant form starting from the uniform energy-density gauge, he needed to introduce 
a redefinition of the first-order perturbations, hence changing their ground state. This redefinition 
corresponds to part of the second-order gauge-invariant curvature perturbation and contributes to the local non-Gaussianity. 
His work was followed by \cite{Chen:2006nt} introducing general kinetic terms, \cite{Seery:2005gb} for two fields and 
\cite{Langlois:2008qf,Arroja:2008yy,Gao:2008dt} for multiple-field models with a generalized kinetic term. 
In \cite{Seery:2005gb} the treatment of two fields in the flat gauge showed that no field redefinitions occur (see also \cite{Rigopoulos:2011eq}). 
Nevertheless, the absence of redefinitions in this case does not mean the absence of local non-Gaussianity, because the action 
was computed in terms of the scalar fields and not in terms of the adiabatic and isocurvature perturbations. This means that in the method of 
\cite{Seery:2005gb} the $\delta N$ formalism or the long-wavelength formalism is needed to compute the final non-Gaussianity, which requires 
that the slow-roll approximation is imposed at horizon crossing.

Here we generalize the above results and write the two-field action in terms of the gauge-invariant perturbations themselves, 
both in the uniform energy-density gauge and the flat gauge in order to compare the results.
We expand the calculation to include second-order perturbations and tensor modes, and study the 
various contributions that occur. Hence we derive the exact third-order action, going beyond the slow-roll or the 
super-horizon approximation. We thus provide the missing tool that will enable people to calculate non-Gaussianity, using the in-in 
formalism \cite{Weinberg:2005vy}, beyond these standard approximations used by both the long-wavelength formalism and the $\delta N$ formalism.

This paper is organized as follows: in section \ref{back} we provide the gauge-invariant definitions and conventions for the metric 
and the field perturbations, along with the description of the space-time of the universe, using the ADM formalism. 
In the first part of section \ref{gic} we study the gauge-invariant curvature and isocurvature perturbations in terms of the fields, 
while next, in subsection \ref{grad}, we make the connection to the fully non-linear spatial gradients of the relevant quantities. 
In the whole of section \ref{gic} we use the long-wavelength approximation to keep the calculations short and tractable, but we present 
the generalization of the results beyond this approximation in appendix \ref{app0}.  
In section 
\ref{cubic} we construct the exact cubic action, going beyond the long-wavelength approximation, to find the redefinitions of the 
perturbations and compare their contributions to the gauge-invariant 
quantities found in subsection \ref{gaugetrans}. To keep the main text more accessible, many of the details of the 
calculations have been moved to the appendices. 
Finally, in section \ref{summary} we summarize the results from section \ref{cubic}, and we conclude in section \ref{concl}.

\section{Preliminaries}\label{back}

In this section we give the basic elements required for the calculations in this paper. We start by 
summarizing in subsection \ref{adm} the ADM formalism, along with the definitions of the cosmological 
quantities, the slow-roll parameters and the field basis we use. In subsection \ref{def} we provide the conventions of cosmological perturbation theory and clarify 
different approaches in the literature.

\subsection{The ADM formalism}\label{adm}

We will consider a universe filled with two scalar fields with a trivial
field metric. The generalization to more fields and a non-trivial
field metric is conceptually straightforward (see \cite{Rigopoulos:2005xx}), but involves more complicated 
expressions and calculations.   
 The energy-momentum tensor for the two fields $\varphi^A$ ($A,B = 1,2$) is
\be
T_{\mu\nu}=\gd_{AB}\partial_{\mu}\varphi^A\partial_{\nu}\varphi^B-g_{\mu\nu}
\left(\frac{1}{2}\gd_{AB}g^{\lambda\kappa}\partial_{\kappa}\varphi^A\partial_{\lambda}\varphi^B
+W\right),
\ee
where $W$ is the field potential. We will denote the homogeneous part of the fields by $\phi^A$. 
The Einstein summation convention is assumed
throughout this paper.
We shall work in the ADM formalism and write the metric $g_{\mu\nu}$ as
\be
ds^2=-\bN^2dt^2+h_{ij}(dx^i+N^idt)(dx^j+N^jdt),\label{metricexact}
\ee
where $\bN$ is the lapse function and $N^i$ the shift. The action takes the form \cite{Misner:1974qy}
\be
S=\frac{1}{2}\int\d^4x\sqrt{h}\Big[-2\bN W+\kappa^{-2}\bN^{-1}(E_{ij}E^{ij}-E^2)+\bN \bar{\Pi}^2
+\kappa^{-2}\bN R^{(3)}-\bN h^{ij}\partial_i\varphi^A\partial_j\varphi_A\Big],\label{actionexact}
\ee
where $\kappa^2\equiv8\pi G=8\pi/m_{pl}^2$, $h$ is the determinant of the space metric $h_{ij}$, $R^{(3)}$ is the intrinsic 
3-curvature, the tensor $E_{ij}$ (proportional to the extrinsic curvature $K_{ij}=-\bN^{-1}E_{ij}$) is
\be
E_{ij}=\frac{1}{2}\left(\dot{h}_{ij}-\nabla_iN_j-\nabla_jN_i\right)
\ee
and $\bar{\Pi}$ is the length of the canonical momentum of the fields 
\be
\bar{\Pi}^A=(\dvphi^A-N^i\partial_i\varphi^A)/\bN.
\ee
Variation of the action with respect to $\bar{N}$ and $N^i$ gives the energy and momentum constraints
\bea
&&\kappa^{-2}R^{(3)}-2W-\kappa^{-2}\bN^{-2}(E_{ij}E^{ij}-E^2)-\bar{\Pi}^2-h^{ij}\partial_i\varphi^A\partial_j\varphi_A=0,\label{energy}\\
&&\nabla_j\Big[\frac{1}{\bN}(E_i^j-E\delta_i^j)\Big]=\kappa^2\bar{\Pi}^A\partial_i\varphi_A,\label{momentum}
\eea
where $\nabla_j$ denotes the covariant derivative with respect to the space metric and $E$ is the trace of $E_{ij}$.

Following \cite{Maldacena:2002vr} we decompose the space metric as
\be
h_{ij}=a(t)^2e^{2\alpha(t,x)}e^{\gamma_{ij}(t,x)},\qquad \partial_i\gamma^{ij}=0,\qquad\gamma_i^i=0.\label{metric}
\ee
From now on contravariant tensors should be understood as $T^j=\eta^{ij}T_i$, where $\eta^{ij}=\mathrm{diag}(1,1,1)$, since 
in the calculations we are showing we have already taken into account explicitly the $h^{ij}$ part of the initial 
contravariant tensors. 
The generalized Hubble parameter is defined as 
\be
\bar{H}\equiv\frac{E}{3\bar{N}}.
\ee 
We use the bar for the lapse function, the Hubble parameter and the canonical momentum to distinguish these fully non-linear quantities 
from their background values $N(t)$, $H(t)=\dot{a}/(a N)$ and $\Pi(t)=\dphi/N$, respectively. 
In this paper we will use as time variable the number of e-folds, meaning that $\dot{a}=a$, so that the background 
value of the lapse function $N(t)$ is just $1/H(t)$.

The background field equation and the background Einstein equations are
\bea
\dot{\Pi}^A= -3HN\Pi^A-N W^{A},\qquad \dot{H}=-\frac{\kappa^2}{2}N\Pi^2,\qquad H^2=\frac{\kappa^2}{3}\lh\frac{\Pi^2}{2}+W\rh,
\label{fieldeq}
\eea
with $W_{A}\equiv\der W/\der\gf^A$. 
We construct an orthonormal basis ${e_m^A}$ in field space, consisting of $e_1^A\equiv\Pi^A/\Pi$, parallel to the field velocity, 
and $e_2^A$ parallel to the part of the field acceleration perpendicular to the field velocity  
 \cite{GrootNibbelink:2001qt}. The $m=1$ component of physical quantities describes the single-field (adiabatic) part, while 
the $m=2$ component captures the multiple-field (isocurvature) effects. 
One can show that for the two-field case the basis vectors are related through 
\cite{Tzavara:2010ge}
\be
\epsilon_{AB}e_1^Ae_2^B=-1,
\label{antisym}
\ee 
where $\e^{AB}$ is the antisymmetric tensor. The background slow-roll parameters then take the form 
\bea
\ge(t)  &\equiv& - \frac{\dot{H}}{NH^2},
\quad
\hpa(t) \equiv \frac{e_{1A}\dot{\Pi}^A}{NH\Pi},
\quad 
\hpe(t) \equiv \frac{e_{2A}\dot{\Pi}^A}{NH\Pi},\quad
\chi(t) \equiv \frac{W_{22}}{3H^2}+\e+\hpa,\non\\
\gx^\parallel(t) &\equiv& \frac{e_{1A}\ddot{\Pi}^A}{N^2H^2\Pi}-\frac{\dot{N}}{N^2H}\hpa,
\qquad\!\!
\gx^\perp(t) \equiv \frac{e_{2A}\ddot{\Pi}^A}{N^2H^2\Pi}-\frac{\dot{N}}{N^2H}\hpe,
\label{srvar}
\eea
where $W_{mn}\equiv e_m^Ae_n^BW_{,AB}$. 
Throughout this paper the indices $m, n$ will indicate components in the
basis defined above, taking the values 1 and 2, while $i,j$ are spatial indices and $A,B$ are indices of the original fields. 
In a slow-roll approximation one can think of $\hpa$ being related to $W_{11}$, $\hpe$ to $W_{21}$ and 
$\chi$ to $W_{22}$.
The $\xi$ parameters are second-order slow-roll parameters (in a slow-roll approximation they are related to 
the third derivatives of the potential). However, we emphasize that we have not made any slow-roll approximations; the above 
quantities should be viewed as short-hand notation and can be large.
We also give the time derivatives of the background slow-roll parameters and of the unit vectors,
\bea
&&\dot{\ge} = 2 NH \ge ( \ge + \get^\parallel ),\quad
\dot{\get}^\parallel\!= NH\!\lh\gx^\parallel \!+ (\get^\perp)^2 + (\ge - \get^\parallel) \get^\parallel\rh, \quad
\dot{\get}^\perp\! =NH\!\lh \gx^\perp \!+ (\ge - 2\get^\parallel) \get^\perp\rh,\nn\\
&&\dot{\gc} =NH\lh \ge \get^\parallel + 2 \ge \gc - (\get^\parallel)^2
+ 3 (\get^\perp)^2 + \gx^\parallel + \frac{2}{3} \get^\perp \gx^\perp
+ \frac{\sqrt{2\ge}}{\gk}\frac{W_{221}}{3H^2}\rh,\nn\\
&&\dot{\xi}^\parallel=NH\lh -\frac{\sqrt{2\ge}}{\gk}\frac{W_{111}}{H^2}+2\hpe\gx^\perp+2\e\gx^\parallel-3\gx^\parallel+9\e\hpa
+3(\hpe)^2+3(\hpa)^2\rh,\nn\\
&&\dot{\xi}^\perp=NH\lh -\frac{\sqrt{2\ge}}{\gk}\frac{W_{211}}{H^2}-\hpe\gx^\parallel+2\e\gx^\perp-3\gx^\perp+9\e\hpe+6\hpe\hpa-3\hpe\chi\rh,\nn\\
&&\dot{e}_1^A= NH\hpe e_2^A,\qquad\qquad \dot{e}_2^A=- NH\hpe e_1^A,\label{dere}
\eea
where $W_{lmn}\equiv e_m^Ae_n^Be_l^CW_{,ABC}$.

\subsection{Second-order perturbations and gauge transformations}\label{def}

In the context of perturbation theory around an homogeneous background any quantity 
$\bar{A}$ can be 
decomposed into an homogeneous part and an infinite series of perturbations as 
\be
\bar{A}(t,\vc{x})=A(t)+A_{(1)}(t,\vc{x})+\frac{1}{2}A_{(2)}(t,\vc{x})+\dots, 
\ee
where the subscripts in the parentheses denote the order of the perturbation. Up to first order the scalar part of the 
space metric element of (\ref{metric}) is equal to
\be
h_{ij}=a^2(t)(1+2\ae)\delta_{ij}.\label{first}
\ee
When one wants to expand perturbation theory up to second order there are two choices found in the literature: either expand (\ref{first}) 
as Malik and Wands do in \cite{Malik:2003mv} to find
\be
h_{ij}=a^2(t)(1+2\ae+\aee)\delta_{ij} 
\ee
or expand directly the space part of (\ref{metric}) as for example Lyth and Rodriguez do in 
\cite{Lyth:2005du} to find
\be
h_{ij}=a^2(t)(1+2\ae+2\ae^2+\aee)\delta_{ij}.\label{second}
\ee
We will take this second approach and use the exponent of the perturbation in our calculations.

Since perturbations depend on the gauge choice we make, we need to construct quantities that are invariant 
under gauge transformations. Under an arbitrary second-order coordinate transformation
\be
\widetilde{x}^{\mu}=\hat{x}^{\mu}+\beta_{(1)}^{\mu}+\frac{1}{2}\lh\beta_{(1),\nu}^{\mu}\beta^{\nu}_{(1)}+\beta_{(2)}^{\mu}\rh,
\ee
the perturbations of a tensor transform as \cite{Bruni:1996im}
\bea
&&\widetilde{A}_{(1)}=\hat{A}_{(1)}+L_{\beta_{(1)}}A,\nn\\
&&\widetilde{A}_{(2)}=\hat{A}_{(2)}+L_{\beta_{(2)}}A+L_{\beta_{(1)}}^2A+2L_{\beta_{(1)}}\hat{A}_{(1)},\label{trans}
\eea
where $L_{\beta}$ is the Lie derivative along the vector $\beta$
\be
\lh L_{\beta}A\rh^{\mu_1\mu_2...}_{\nu_1\nu_2...}=\beta^\kappa\partial_\kappa A^{\mu_1\mu_2...}_{\nu_1\nu_2...}
-\partial_\kappa\beta^{\mu_1}
A^{\kappa\mu_2...}_{\nu_1\nu_2...}-\dots +\partial_{\nu_1}\beta^{\kappa}A^{\mu_1\mu_2...}_{\kappa\nu_2...}+\dots .
\ee
Note here that spatial gradients, having vanishing background 
values, are automatically gauge-invariant at first order, while at second order they transform as
\be
\partial_i\widetilde{A}_{(2)}=\partial_i\hat{A}_{(2)}+2L_{\beta_{(1)}}\partial_i\hat{A}_{(1)}.\label{zerotr}
\ee

\section{Super-horizon gauge transformations}\label{gic}

In this section we derive first and second-order super-horizon gauge-invariant combinations.  
We study these during an inflationary period and thus, though we start 
from the energy-density definitions of the perturbations, we naturally end up with field definitions for the gauge-invariant 
perturbations. 
Our goal is to find the second-order adiabatic and isocurvature perturbations in terms of the first-order ones and 
the slow-roll parameters. 

In this section we restrict ourselves to super-horizon calculations for simplicity, though in the next section we will 
abandon this approximation and study the full action of the cosmological perturbations. 
However in appendix \ref{app0} we present the generalization of the results of this section 
beyond the long-wavelength 
approximation. We note that the long-wavelength (or super-horizon) approximation 
is equivalent to the zeroth order space gradient approximation and is valid once the decaying mode has disappeared (which happens 
rapidly if slow-roll holds during horizon exit), even if there is a subsequent non slow-roll phase.

In the super-horizon regime one can choose to work in the time-orthogonal gauge where $N^i=0$ 
(proof for that choice is given in the next section) and employ the long-wavelength approximation to simplify calculations. 
The latter boils down to ignoring second-order spatial derivatives when compared to time derivatives. 
As a consequence the traceless part of the extrinsic curvature quickly decays and 
can be neglected \cite{Salopek:1990re}. Hence the space part of the metric can be described by 
\be
h_{ij}=a(t)^2e^{2\alpha(t,x)}\delta_{ij}.
\ee
The field and Einstein equations in that case are identical to (\ref{fieldeq}), but now the quantities involved are fully non-linear. 
Additionally the momentum constraint (\ref{momentum}) can be written as \cite{Rigopoulos:2005xx}
\be
\partial_i\bar{H}=-\frac{\kappa^2}{2}\bar{\Pi}_A\partial_i\varphi^A.\label{fieldeqsu}
\ee 

\subsection{Gauge-invariant quantities}\label{gaugetrans}

The well-known first-order adiabatic gauge-invariant curvature perturbation has the form 
\be
\zeta_{1(1)}\equiv\ae-\frac{NH}{\dot{\rho}}\rho_{(1)},\label{z110}
\ee
where $\rho$ is the energy density. The subscript without parentheses corresponds to the first component in our basis, 
which is exactly the adiabatic component, while the 
subscript between parentheses denotes the order in the perturbation series. 
Notice that in the literature it is common to work with cosmic time, i.e. $N=1$, while the space part of the metric is 
decomposed using a quantity $\psi=-\alpha$ (not to be confused with the $\psi$ introduced in appendix \ref{app2}), 
so that the first-order curvature perturbation becomes in that case 
$-\zeta_{1(1)}=\psi_{(1)}+(H/\dot{\rho})\rho_{(1)}$.
Here we choose to work with the number of e-folds as time variable so that the first-order curvature perturbation is
\be
\zeta_{1(1)}=\ae-\frac{\rho_{(1)}}{\dot{\rho}}.\label{z11}
\ee 

The gauge-invariant combination (\ref{z110}) is calculated via the requirement that it coincides with the curvature 
perturbation $\zeta_{1(1)}=\tilde{\alpha}_{(1)}$ in the uniform energy-density gauge where $\tilde{\rho}_{(1)}=0$. 
 From now on tilded quantities 
will denote the uniform energy-density gauge, 
while hatted quantities will denote the flat gauge. One has to use the gauge 
transformations (\ref{trans}) for a scalar (here the energy density and the logarithm of the space dependent scale factor 
$\alpha$) and require that in the 
uniform energy-density gauge the first-order energy perturbation is zero (for details see appendix \ref{app0}). That way one can 
determine the first-order time  
shift and hence find the gauge-invariant combination corresponding to the curvature perturbation. Notice 
that in the flat gauge, i.e. $\hat{\alpha}_{(1)}=0$, $\zeta_{1(1)}=-\hat{\rho}_{(1)}/\dot{\rho}$. 

Keeping in mind the expansion (\ref{second}), one can repeat the above considerations at second order. 
We find that for super-horizon scales (where we neglect second-order space derivatives when compared to second-order time derivatives) 
the second-order gauge-invariant adiabatic perturbation takes the form (see appendix \ref{app0})
\bea
\frac{1}{2}\zeta_{1(2)}\equiv\frac{1}{2}\tilde{\alpha}_{(2)}
&=&\frac{1}{2}\alpha_{(2)}-\frac{1}{2}\frac{\rho_{(2)}}{\dot{\rho}}
+\frac{\dot{\rho}_{(1)}\rho_{(1)}}{\dot{\rho}^2}
-\frac{\rho_{(1)}}{\dot{\rho}}\dot{\alpha}_{(1)}
-\frac{1}{2}\frac{\rho_{(1)}^2}{\dot{\rho}^2} \frac{\ddot{\rho}}{\dot{\rho}}.\label{defmalik}
\eea
If we chose the second gauge to be flat, i.e. $\ha_{(i)}=0$, we find
\be
\frac{1}{2}\zeta_{1(2)}=\frac{1}{2}\tilde{\alpha}_{(2)}=-\frac{1}{2}\frac{\hr_{(2)}}{\dot{\rho}}
+\dot{\zeta}_{1(1)}\zeta_{1(1)}+\frac{1}{2}\frac{\hr_{(1)}^2}{\dot{\rho}^2} \frac{\ddot{\rho}}{\dot{\rho}}.\label{36}
\ee

During inflation we find it more useful to work directly with the fields and not their energy density, since both 
the long-wavelength formalism and $\delta N$ formalism make use of the field values to compute $f_{NL}$. Using the fields, 
the first-order adiabatic perturbation becomes
\be
\zeta_{1(1)}=\tilde{\alpha}_{(1)}=\alpha_{(1)}-\frac{H}{\Pi}e_{1A}\varphi^A_{(1)},\label{38}
\ee
since the energy-density constraint $\tilde{\rho}_{(1)}=0$ is equivalent to $e_{1A}\tvarphi^A_{(1)}=0$. The detailed calculation 
is shown in appendix \ref{app1}. Notice that in (\ref{38}) we have kept the lapse 
function $N$ arbitrary, as we will also do in all definitions hereafter, but in our calculations 
$H/\Pi$ is just $1/\dphi$ for the choice $N=1/H$.

The second-order calculation turns out to be more complicated. The details are given in appendix \ref{app1}. Here we give the result for 
the gauge 
invariant adiabatic perturbation in the uniform energy-density gauge and in the flat gauge:
\bea
\frac{1}{2}\zeta_{1(2)}=\frac{1}{2}\tilde{\alpha}_{(2)}&=&\frac{1}{2}\hat{Q}_{1(2)}
+\frac{\e+\hpa}{2}\lh\zeta_{1(1)}^2-\zeta_{2(1)}^2\rh-\hpe\zeta_{1(1)}\zeta_{2(1)}+\dot{\zeta}_{1(1)}\zeta_{1(1)}\nn\\
&-&\partial^{-2}\partial^i\lh\dot{\zeta}_{2(1)}\partial_i\zeta_{2(1)}\rh,\label{z1ft}
\eea
where we introduced the auxiliary quantities
\be
Q_{m(i)}\equiv-\frac{H}{\Pi}e_{mA}\varphi^A_{(i)}\
\ee
and the new combination \cite{Rigopoulos:2005xx}
\be
\zeta_{2(1)}\equiv-\frac{H}{\Pi}e_{2A}\varphi^A_{(1)}=Q_{2(1)},\label{z21}
\ee
that represents the isocurvature perturbation to first order. 
Usually the isocurvature perturbation is described in terms of the gradient of the pressure of the matter content of the universe, 
as for example in \cite{Langlois:2005qp}. Here we choose to characterize it in terms of the fields themselves and the vector $e_{2A}$. 
The latter indicates we are dealing with a purely multiple-field effect and hence it is an appropriate quantity to use 
during the inflationary 
period to describe the non adiabatic perturbations. Starting from the long-wavelength definition of the pressure 
$\bar{p}=\bar{\Pi}^2/2-W$, one can show that the gradient of the isocurvature perturbation defined in \cite{Langlois:2005qp} is equal to
\be
\Gamma_i\equiv\partial_i\bar{p}-\frac{\dot{\bar{p}}}{\dot{\bar{\rho}}}\partial_i\bar{\rho}=-2\bar{\eta}^{\perp}\bar{\Pi}^2\zeta_{2i},
\ee
where $\zeta_{2i}$ is the fully non-linear gradient of the isocurvature perturbation 
(for more details see the next subsection \ref{grad}) and $\bar{\eta}^{\perp}$ the fully non-linear generalization 
of $\hpe$, i.e. as it is defined in (\ref{srvar}) but with barred quantities \cite{Rigopoulos:2005xx}. 
So our definition of $\zeta_2$ agrees with the pressure definition of the isocurvature perturbation. 
The next logical step would be to define the second-order isocurvature perturbation $\zeta_{2(2)}$. However, 
the above equation 
shows that there is a non-trivial relation between $\zeta_2$ and the pressure $p$, involving the non-linear quantities 
$\bar{\eta}^{\perp}$ and $\bar{\Pi}$, 
which makes a derivation using the methods of this subsection rather complicated. 
For that reason we prefer to find $\zeta_{2(2)}$ in an easier way in the next section 
using gradients.

Notice that unlike in the original definition of $\zeta_{1(2)}$ in terms of $\rho_{(2)}$, a non-local term appears in (\ref{z1ft}) 
when one uses the fields 
instead of the energy density, because of (\ref{alp_flat}). The time derivatives of the 
fully non-linear gradients of the perturbations (see next section) were found in \cite{Rigopoulos:2005xx}. Expanding  
to first order these yield
\be
\dot{\zeta}_{1(1)}=2\hpe \zeta_{2(1)}
\ee   
for the adiabatic perturbation and
\be 
\dot{\zeta}_{2(1)}=-\chi \zeta_{2(1)}
\ee    
for the isocurvature perturbation, the latter valid only in the slow-roll regime.
Then we find
\bea
\frac{1}{2}\zeta_{1(2)}=\frac{1}{2}\tilde{\alpha}_{(2)}&=&\frac{1}{2}\hat{Q}_{1(2)}
+\frac{\e+\hpa}{2}\zeta_{1(1)}^2-\frac{\e+\hpa-\chi}{2}\zeta_{2(1)}^2+\hpe\zeta_{1(1)}\zeta_{2(1)},
\eea
i.e. without a non-local term.
However, we will not use the slow-roll approximation in this paper.

\subsection{The gradient of the perturbations}\label{grad}

As an alternative to the $\zeta_m$ defined in the previous section, one can use the gradient quantity $\zeta_{1i}$ 
along with the isocurvature analogue $\zeta_{2i}$, both defined in \cite{Rigopoulos:2005xx} and later in 
\cite{Langlois:2005qp} in a covariant way, to construct a gauge-invariant quantity.  
These gradient quantities (not gauge-invariant to all orders) are given by
\be
\zeta_{mi}=\delta_{m1}\partial_i\alpha-\frac{\bar{H}}{\bar{\Pi}}\bar{e}_{mA}\partial_i\varphi^A\label{defgrad},
\ee
where now $\bar{e}_{mA}$ represents the fully non-linear super-horizon version of the orthonormal basis vectors, e.g.  
$\bar{e}_{1A}=\bar{\Pi}_A/\bar{\Pi}$, with $\bar{\Pi}_A=\dot{\varphi}_A/\bar{N}$ since we are working in the super-horizon regime. 
Notice that the basis vectors still obey (\ref{antisym}) as was shown in 
\cite{Tzavara:2010ge}. 
$\zeta_{mi}$ is by construction gauge-invariant at first order, since it has no background value: it is just the gradient of 
the gauge-invariant $\zeta_{m(1)}$ defined before.

Expanding to second order we find for the adiabatic perturbation
\be
\frac{1}{2}\zeta_{1i(2)}=\frac{1}{2}\partial_i\lh\alpha_{(2)}+Q_{1(2)}\rh
-\frac{1}{\dphi^2}\dvphi^A_{(1)}\partial_i\varphi_{A(1)}
-2\frac{1}{\dphi}e_{1A}\dvphi^A_{(1)} \partial_iQ_{1(1)}.\label{akoma}
\ee
In the uniform energy-density gauge this gives (see appendix \ref{app1})
\be
\frac{1}{2}\tilde{\zeta}_{1i(2)}=\frac{1}{2}\partial_i\tilde{\alpha}_{(2)},\label{zai}
\ee
while in the flat gauge where $\partial_i\ha=0$ and $\zeta_{1(1)}=\hat{Q}_{1(1)}$, we find 
\bea
\frac{1}{2}\hat{\zeta}_{1i(2)}
&&=\partial_i\Bigg[\frac{1}{2}\hat{Q}_{1(2)}+\frac{\e+\hpa}{2}\lh\zeta_{1(1)}^2-\zeta_{2(1)}^2\rh-\hpe\zeta_{1(1)}\zeta_{2(1)}
+\dot{\zeta}_{1(1)}\zeta_{1(1)}\Bigg]-\dot{\zeta}_{2(1)}\partial_i\zeta_{2(1)}\nn\\
&&\qquad-\zeta_{1(1)}\partial_i\dot{\zeta}_{1(1)},
\label{z1fg}
\eea
where we used the basis completeness relation and (\ref{difzflat}) to rewrite the terms. 
$\tilde{\zeta}_{1i(2)}$ in the uniform energy-density gauge (\ref{zai}) coincides with the gradient of the 
gauge-invariant second-order adiabatic perturbation. 
However, by comparing (\ref{z1ft}) to (\ref{z1fg}) we see 
that in the flat gauge $\hat{\zeta}_{1i(2)}$ is the gradient 
of the gauge-invariant curvature perturbation $\zeta_{1(2)}$ expressed in the flat gauge plus a new non-local term. 
This is in agreement 
with the findings in \cite{Langlois:2005qp}. This new term is nothing else but the
gauge transformation of $\zeta_{1i(2)}$. A quantity with zero background value as $\zeta_i$ is, transforms as (\ref{zerotr}). 
One can check, using the gauge transformations (\ref{trans}) for $\rho$ and for $\alpha$ 
and requiring that $\tilde{\rho}_{(1)}=0$, 
that starting from a flat gauge and transforming to the uniform energy-density gauge the time shift is 
$T_{(1)}=\zeta_{1(1)}$ (see appendix \ref{app0}), so that
\be
\frac{1}{2}\tilde{\zeta}_{mi(2)}=\frac{1}{2}\hat{\zeta}_{mi(2)}+\zeta_{1(1)}\dot{\zeta}_{mi(1)}.\label{gat}
\ee

Next we try to find the second-order gauge-invariant part of the isocurvature perturbation $\zeta_{2i}$ by expanding (\ref{defgrad}),
\be
\frac{1}{2}\zeta_{2i(2)}=\frac{1}{2}\partial_iQ_{2(2)}
+\frac{1}{\dphi^2}\e^{BA}\dvphi_{B(1)}\partial_i\varphi_{A(1)}-2\frac{1}{\dphi}e_{1A}\dvphi^A_{(1)} \partial_i\zeta_{2(1)},
\ee
or equivalently,
\be
\frac{1}{2}\zeta_{2i(2)}=\frac{1}{2}\partial_iQ_{2(2)}
-\frac{1}{\dphi}e_{2A}\dvphi^A_{(1)} \partial_iQ_{1(1)}-\frac{1}{\dphi}e_{1A}\dvphi^A_{(1)} \partial_i\zeta_{2(1)},\label{another}
\ee
where we used (\ref{antisym}) to express $\vc{e}_{2}$ in terms of $\vc{e}_1$. 
The first-order uniform energy constraint $\partial_i\tilde{Q}_{1(1)}=0$ alone implies
\bea
\frac{1}{2}\tilde{\zeta}_{2i(2)}&=&\partial_i\Bigg[\frac{1}{2}\tilde{Q}_{2(2)}-\frac{\hpe}{2}\zeta_{2(1)}^2\Bigg]
+\dot{\zeta}_{1(1)}\partial_i\zeta_{2(1)}\label{z2u},
\eea
where we used (\ref{difzue}) for the two last terms in (\ref{another}). 
For the flat gauge $\partial_i\ha=0$ we find using (\ref{difzflat})
\bea
\frac{1}{2}\hat{\zeta}_{2i(2)}&=&\partial_i\Bigg[\frac{1}{2}\hat{Q}_{2(2)}
+\frac{\hpe}{2}\lh\zeta_{1(1)}^2-\zeta_{2(1)}^2\rh
+(\e+\hpa)\zeta_{1(1)}\zeta_{2(1)}+\dot{\zeta}_{2(1)}\zeta_{1(1)}\Bigg]+\dot{\zeta}_{1(1)}\partial_i\zeta_{2(1)}\nn\\
&&-\zeta_{1(1)}\partial_i\dot{\zeta}_{2(1)}\label{z2f}.
\eea
We notice that the term in the second line corresponds again to a gauge transformation familiar from the curvature perturbation case 
studied earlier (\ref{gat}). In appendix \ref{app0} we verify that the rest of the expression is a gauge-invariant quantity 
corresponding to the one in (\ref{z2u}). Indeed this expression is gauge-invariant beyond the long-wavelength approximation 
as shown in appendix \ref{app0}.

We conclude that the gradients of the perturbations are in some sense equivalent to the perturbations themselves, since both 
allow for the definition  of gauge-invariant second-order adiabatic and isocurvature quantities. However, since the 
gradients are defined using fully non-linear quantities, their equations of motion can be treated 
more easily, as was shown in \cite{Rigopoulos:2005xx}.

\section{The cubic action}\label{cubic}
\label{appSecondSource}

An alternative way to calculate the second-order gauge-invariant quantities and reconsider their meaning, is to compute the third-order 
action for the interacting fields. 
Maldacena \cite{Maldacena:2002vr} was the first to perform that calculation for a single field, in the uniform 
energy-density gauge. 
In this way he managed to find the cubic interaction terms due to non-linearities of the Einstein action as well as 
of the field potential, which among 
other consequences change the ground state of the adiabatic perturbation $\zeta_{1(1)}$. 
This change can be quantified through a redefinition of the 
form \cite{Maldacena:2002vr}
\be
\zeta_{1(1)}=\zeta_{1c(1)}+\frac{\e+\hpa}{2}\zeta_{1(1)}^2,
\ee
where $\zeta_{1c}$ is the redefined perturbation. One sees that the correction term of the redefinition coincides with the 
surviving quadratic term of the single-field limit of the transformation (\ref{z1ft}), taking into account 
that the super-horizon adiabatic perturbation is constant in that case.
In \cite{Maldacena:2002vr} the curvature perturbation was considered a first-order quantity, 
while the second-order curvature perturbation was not taken into account, since its contribution in the uniform 
energy-density gauge 
is trivial: it introduces a redefinition of the form $\zeta_{1(1)}+\zeta_{1(2)}/2=\zeta_{1c(1)}$ 
(for proof, see subsection \ref{cub}). 
Seery and Lidsey \cite{Seery:2005gb} performed the same calculation for the multiple-field case in the flat gauge 
in terms of the scalar fields $\varphi^A$ and not of the adiabatic and isocurvature perturbations $\zeta_m$. They found 
no redefinitions, but as mentioned before their results would have to be supplemented 
by the $\delta N$ formalism (with its associated slow-roll approximation at horizon-crossing) to say anything about the non-Gaussianity 
of the gauge-invariant perturbations $\zeta_m$.

In this section we generalize the above calculations to second order in the expansion of the curvature perturbation in both the uniform 
energy-density gauge and the flat gauge. Doing so we compute the full form of the third-order action. The latter not only 
consists of the cubic interactions of the first-order curvature perturbations, but also of lower order interaction terms of the second 
order quantities. We first perform the calculation relevant to the first-order quantities and then add the second-order 
effects. In this section we only present the scalar part of the action, but in appendices \ref{app2} and \ref{app3} the 
tensor part can be found as well. We emphasize that in this section we no longer make the long-wavelength approximation, so that 
the results are valid at any scale.

\subsection{The second-order action}

We start by performing our calculation in the gauge $e_{1A}\tilde{\varphi}^A_{(1)}=0$. This constraint reduces to the 
uniform energy-density 
gauge outside the horizon, which is why we will continue to refer to the tilded gauge as the uniform energy-density gauge. 
From now on we drop the explicit 
subscript $(1)$ on first-order quantities. We will keep this part brief since its results are 
already known, but we give the basic elements of the calculation in appendix \ref{app2}. 
The second-order action takes the form 
\bea
\tilde{S}_2&=&\!\!\int\!\! \d^4x L_2\nn\\
&=&\!\!\int\!\! \d^4x\ \e \Bigg\{\!\!-a\frac{1}{H}\Big( (\partial\zeta_1)^2+(\partial\zeta_2)^2\Big)
 +a^3H\Big(\dot{\zeta}_1^2+\dot{\zeta}_2^2-4\hpe \dot{\zeta}_1\zeta_2
+2\chi \dot{\zeta}_2\zeta_2\Big)\label{s2}\\
&&\qquad\quad+a^3H\Big(\!\sqrt{\frac{2\e}{\kappa}}\frac{W_{221}}{3H^2}-2\e^2-(\hpa)^2+3(\hpe)^2+\frac{2}{3}\hpe\gx^\perp
-3\e(\hpa-\chi)+2\hpa\chi\Big) \zeta_2^2\Bigg\}\nn,
\eea
where $L_2$ is the second-order Lagrangian. 
While we have started from an action describing the evolution of the fields $\tilde{\alpha}$ and $e_{2A}\tvarphi^A$ we have now 
constructed an action 
in terms of the adiabatic and isocurvature perturbations $\zeta_1$ and $\zeta_2$.
The equations of motion that $\zeta_1$ and $\zeta_2$ obey are ($\delta L_2/\delta\zeta_m$ being a short-hand notation for the 
relevant variations of the Lagrangian)
\bea
&&\frac{\delta L_2}{\delta\zeta_1}=-2a^3\e H\Big[\ddot{\zeta}_1+\lh3+\e+2\hpa\rh \dot{\zeta}_1-2\hpe \dot{\zeta}_2
+2(-\gx^\perp-2\e\hpe-3\hpe)\zeta_2\Big]+2a\frac{\e}{H}\partial^2\zeta_1\nn\\
&&\qquad=-\frac{\d}{\d t}(2a^3H\partial^2\lambda)+2a\frac{\e}{H}\partial^2\zeta_1=0,\label{eq}\\
&&\frac{\delta L_2}{\delta\zeta_2}=-2a^3\e H\Big[\ddot{\zeta}_2+\!\lh3+\e+2\hpa\rh \dot{\zeta}_2+2\hpe \dot{\zeta}_1+(\gx^\parallel+2\e^2+4\e\hpa+3\chi)\zeta_2
\Big]\!+\!2a\frac{\e}{H}\partial^2\zeta_2=0,\nn 
\eea
where $\partial^2\lambda=\e\dot{\zeta}_1-2\e\hpe\zeta_2$ (for the reason of introducing $\lambda$ see appendix \ref{app2}).
Thus we have found the evolution equations for the first-order adiabatic and isocurvature perturbations. Their super-horizon limit 
coincides with the equations derived in \cite{Rigopoulos:2005xx} for the gradient of the perturbations, since up to first 
order $\zeta_{mi(1)}=\partial_i\zeta_{m(1)}$. 
One can show that the first-order energy constraint, which outside the horizon reduces to 
\be
\dot{\zeta}_1-2\hpe\zeta_2=0,\label{mc}
\ee
is the first integral of the super-horizon part of the first equation of (\ref{eq}), i.e. without the space gradient. 
In fact it was shown in \cite{Rigopoulos:2005us} that this is the case at all orders. 
In the same paper it was found that assuming the slow-roll limit, $\dot{\zeta}_2=-\chi\zeta_2$ is the super-horizon first integral of 
the equation for $\zeta_2$, which can be easily verified. 

While working in the flat gauge we find the same action (\ref{s2}) (see appendix \ref{app2}). So the curvature perturbations 
$\zeta_m$ 
satisfy to first order the same equations in both gauges as expected, due to the gauge invariance of $\zeta_m$ 
(or equivalently the gauge invariance of the action).

\subsection{The third-order action}\label{cub}

In this section we compute the third-order action. 
Again we present only the final results, while in appendix \ref{app3} we give the intermediate steps of the 
calculation. In the same appendix we also give the tensor-scalar part of the action. 
The scalar cubic action in the uniform energy-density gauge due to the first-order perturbations $\zeta_m$ takes the form
\bea
\tilde{S}_{3(1)}\!&=&S_{3(1)}-\int\d^4x\frac{\delta L_2}{\delta\zeta_m}f_m
\eea
with 
\bea
&&f_1=\frac{\e+\hpa}{2}\zeta_1^2-\hpe\zeta_1\zeta_2
+\dot{\zeta}_1\zeta_1-\frac{1}{4a^2H^2}\Big((\partial\zeta_1)^2
-\partial^{-2}\partial^i\partial^j(\partial_i\zeta_1\partial_j\zeta_1)\Big)
\nn\\
&&\qquad
+\frac{1}{2}\Big(\partial^i\lambda\partial_i\zeta_1-\partial^{-2}
\partial^i\partial^j(\partial_i\lambda\partial_j\zeta_1)\Big),\nn\\
&&f_2=(\e+\hpa)\zeta_1\zeta_2+\dot{\zeta}_2\zeta_1+\frac{\hpe}{2}\zeta_1^2\label{rede1}.
\eea
The exact form of $S_{3(1)}$ can be found in appendix \ref{app3} or equivalently it is the cubic part of (\ref{fin}). 
The reason for introducing $S_{3(1)}$ without the tilde will become clear below. 

The terms proportional to 
$\delta L_2/\delta\zeta_m$, i.e. the first-order equations of motion, 
can be removed  by a redefinition of $\zeta_m$
\cite{Maldacena:2002vr} and lead to a change in the ground state of the perturbations. This works as follows.  
The cubic terms of the action (i.e. $\tilde{S}_{3(1)}$) are not affected by the redefinition, because the redefinition always 
involves terms proportional to $\zeta_m^2$, which would give quartic and not cubic corrections. 
It is only the second-order 
terms (i.e. $\tilde{S}_2$) that change. 
Indeed one can show that under a redefinition of the form $\zeta_m=\zeta_{mc}+f_m$, the second-order action changes as 
$S_2=S_{2c}+(\delta L_2/\delta\zeta_m)f_m$. These new terms cancel out the relevant terms coming from the cubic action 
(remember that the total action up to cubic order is the sum of the second and third-order action) and we are left 
with
\be
\tilde{S}_{3(1)}=S_{3(1)}(\zeta_{mc}).\label{ss3}
\ee

If we repeat the same calculations for the flat gauge (see appendix \ref{app3}), 
performing several integrations by part, we find that 
\be
\hat{S}_{3(1)}=S_{3(1)}(\zeta_{m}).
\ee 
This is a consequence of the action staying invariant under a gauge transformation. 
Nevertheless if one associates the redefinition appearing in the uniform energy-density gauge to a change in the ground state 
of $\zeta_m$, it would mean that directly after horizon crossing, when super-horizon effects have not yet 
been switched on, the second-order contribution to $\zeta_1$ would be zero for the flat gauge and non-zero for the uniform energy-density 
gauge. In terms of non-Gaussianity, this can be restated as: the non-Gaussianity present after horizon-crossing is different for the 
two gauges. 
Indeed if one was to calculate the three-point functions for the above action, one would need to perform 
two steps. First, change to the interaction picture, where it can be proved that the interaction Hamiltonian up to and 
including cubic order is just $H_{int}=-L_{int}$, where $L_{int}$ are the cubic terms of the Lagrangian, and compute the 
expectation value $\langle\zeta_c\zeta_c\zeta_c\rangle$ as in \cite{Maldacena:2002vr}. Second, take into account that the 
fields have been redefined as $\zeta=\zeta_c+\lambda\zeta_c^2$. Then the three-point correlation function can be 
written as
\be
\langle\zeta\zeta\zeta\rangle=\langle\zeta_c\zeta_c\zeta_c\rangle
+2\lambda[\langle\zeta_c\zeta_c\rangle
\langle\zeta_c\zeta_c\rangle+\mathrm{cyclic}].\label{cyclic}
\ee
These new terms, products of the second-order correlation functions, are only present in the uniform energy-density gauge 
if we restrict ourselves to $S_{3(1)}$. 

In order to cure this bad behavior we need to add to the above results the effect of the second-order fields. 
We find (see appendix \ref{app3})
\bea
\tilde{S}_{3(2)}=\int\d^4x\Bigg\{\frac{\delta L_2}{\delta\zeta_1}\lh\frac{\tilde{Q}_{1(2)}}{2}+\frac{\zeta_{1(2)}}{2}\rh+
\frac{\delta L_2}{\delta\zeta_2}\frac{\tilde{Q}_{2(2)}}{2}\Bigg\}.
\eea
Since all terms in $\tilde{S}_{3(2)}$ are proportional to $\delta L_2/\delta\zeta_m$, $\tilde{S}_{3(2)}$ only contains 
redefinitions of $\zeta_m$. 
Notice that the second-order lapse and shift functions do not appear in the final action, since these two 
are multiplied by a factor equal to the energy and momentum constraint equations (\ref{n1ni}). 
On the other hand, the second-order field perturbations are dynamical variables that obey second-order equations of motion that cannot 
be set to zero in the action. The single-field limit of this 
action is just the term proportional to $\zeta_{1(2)}$, since $\tilde{Q}_{m(i)}=0$ identically in that case for 
the uniform energy-density gauge. 
The term proportional to $\zeta_{1(2)}$ in $\tilde{S}_{3(2)}$, along with the terms proportional to $\lambda$ in $\tilde{S}_{3(1)}$, 
originate from the contribution of $N^i$ in the action. The latter vanish outside the horizon since then $\partial^2\lambda$ 
coincides with the super-horizon energy constraint and hence is identically zero. 
So if we were to study only the quadratic contributions of the first-order perturbations outside the horizon, 
we would be allowed not only to ignore the tensor parts of the metric \cite{Salopek:1990re}, 
but also work in the time-orthogonal gauge $N^i=0$.
 
Coming back to the redefinition, its final form, including the tensor parts (see appendix \ref{app3}), is
\bea
&&\zeta_1=\zeta_{1c}-\frac{\zeta_{1(2)}}{2}-\frac{\tilde{Q}_{1(2)}}{2}+\dot{\zeta_1}\zeta_1+\frac{\e+\hpa}{2}\zeta_1^2
-\hpe\zeta_1\zeta_2
-\!\frac{1}{4a^2H^2}\Big((\partial\zeta_1)^2
-\partial^{-2}\partial^i\partial^j(\partial_i\zeta_1\partial_j\zeta_1)\Big)\nn\\
&&\qquad
+\frac{1}{2}\Big(\partial^i\lambda\partial_i\zeta_1-\partial^{-2}\partial^i\partial^j(\partial_i\lambda\partial_j\zeta_1)\Big)
-\frac{1}{4}\partial^{-2}(\dot{\gamma}_{ij}\partial^i\partial^j\zeta_1),\nn\\
&&\zeta_2=\zeta_{2c}-\frac{\tilde{Q}_{2(2)}}{2}+\dot{\zeta_2}\zeta_1+\frac{\hpe}{2}\zeta_1^2+(\e+\hpa)\zeta_1\zeta_2.\label{rede2}
\eea

Finally we perform the above calculations for the flat gauge and find the action
\bea
\hat{S}_{3(2)}=\int\d^4x\Bigg\{\frac{\delta L_2}{\delta\zeta_1}\frac{\hat{Q}_{1(2)}}{2}
+\frac{\delta L_2}{\delta\zeta_2}\frac{\hat{Q}_{2(2)}}{2}\Bigg\}.
\eea
The redefinitions in the flat gauge take the simple form
\bea
&&\zeta_1=\zeta_{1c}-\frac{\hat{Q}_{1(2)}}{2}\nn\\
&&\zeta_2=\zeta_{2c}-\frac{\hat{Q}_{2(2)}}{2}.\label{red}
\eea

We want to write these redefinitions as well as the action itself in terms of gauge-invariant quantities and compare them. We would also like 
to compare with the definitions of the second-order gauge-invariant perturbations found in appendix \ref{app0} and section \ref{gic}. 
After using the second-order uniform energy constraint (\ref{con}) and the uniform energy gauge definition of $\zeta_{2(2)}$ 
(\ref{z2u}) we can rewrite (\ref{rede2}) as
\bea
&&\zeta_1+\frac{\zeta_{1(2)}}{2}=\zeta_{1c}+\dot{\zeta_1}\zeta_1+\frac{\e+\hpa}{2}\lh\zeta_1^2-\zeta_2^2\rh-\hpe\zeta_1\zeta_2
-\partial^{-2}\partial^i\lh\dot{\zeta}_2\partial_i\zeta_2\rh
-\frac{1}{4}\partial^{-2}(\dot{\gamma}_{ij}\partial^i\partial^j\zeta_1)
\nn\\
&&\qquad\qquad\quad-\frac{1}{4a^2H^2}\Big((\partial\zeta_1)^2
-\partial^{-2}\partial^i\partial^j(\partial_i\zeta_1\partial_j\zeta_1)\Big)
+\frac{1}{2}\Big(\partial^i\lambda\partial_i\zeta_1-\partial^{-2}\partial^i\partial^j(\partial_i\lambda\partial_j\zeta_1)\Big)\nn\\
&&\zeta_2+\frac{\zeta_{2(2)}}{2}=\zeta_{2c}+\dot{\zeta_2}\zeta_1+\zeta_2\dot{\zeta}_1+\frac{\hpe}{2}\lh\zeta_1^2-\zeta_2^2\rh+(\e+\hpa)\zeta_1\zeta_2
-\partial^{-2}\partial^i\lh\zeta_2\partial_i\dot{\zeta}_1\rh.\label{finu}
\eea
When comparing the first equation of (\ref{finu}) with (\ref{adi}), we see that we recover (\ref{red}). The same is true for the isocurvature 
part of the redefinition: comparing the second equation of (\ref{finu}) with (\ref{z2f}), we recover the redefinition 
for $\zeta_2$ (\ref{red}). 
Hence the two redefinitions are the same, as is necessary for the action to be gauge-invariant. 
Notice that the single-field limit of (\ref{finu}) is 
$\zeta_1+\zeta_{1(2)}/2=\zeta_{1c}+(\e+\hpa)\zeta_1^2/2$ 
in agreement with the total redefinition found in the uniform energy-density gauge. 

Equation (\ref{finu}) is the implicit definition of the redefined, gauge-invariant $\zeta_{mc}$. One can see that up to 
and including second order, it is a function of only the combination $\zeta_{m(1)} + \zeta_{m(2)}/2$. One can also notice 
that the purely second-order perturbation $\zeta_{m(2)}$ does not occur explicitly in the cubic action (see e.g. (\ref{fin}) below). 
Hence once could in principle
consider the quantities $\zeta_{m(2)}$ (and similarly $Q_{m(2)}$) as auxiliary quantities and try to avoid introducing them in the 
first place, but consider the quadratic first-order terms directly as a correction to the first-order perturbations, as is done for 
the single-field case in \cite{Rigopoulos:2011eq}. While the calculations would be roughly equivalent, we have chosen not to follow this route 
for two reasons. In the first place it seems conceptually simpler to us to expand the perturbations and the action consistently up to 
the required order, and more logical to view quadratic first-order terms as a correction to a second-order quantity than to a first-order 
one. Secondly, in the multiple-field case (as opposed to the single-field case), one would have to introduce the second-order quantities at 
some intermediate steps anyway in order to find the correct non-linear relation between the $Q_m$ and $\zeta_m$ (which is derived from the 
second-order gauge transformation performed in appendix \ref{app0}). 

So in the end we have managed to find the source of the non-Gaussianity present at horizon crossing due to first-order perturbations 
and identify it with the quadratic 
terms of (\ref{finu}). With source here we mean the second-order perturbation that, when contracted with two first-order 
perturbations, gives the bispectrum. The super-horizon limit of (\ref{finu}) was derived and used in our previous paper \cite{Tzavara:2010ge}, 
but here we have not only generalized the result, but also have obtained a much better understanding of the gauge issues. 
Equation (\ref{finu}) is gauge-invariant, as it should be. 
Additionally, the redefinition of the perturbations  
that we perform is essential not only to simplify calculations but also to find the gauge-invariant form of the 
action itself. 
We clearly see that the quadratic corrections in the flat gauge seem to be zero if one takes into account only the first-order fields. 
In that gauge all of the second-order contributions are hidden in the second-order fields as opposed to the uniform energy-density 
gauge where part of the quadratic contributions is attributed to the redefinition of the first-order $\zeta_m$ and the rest of them lie 
in the second-order field. 

\subsection{Summary}\label{summary}

Let us summarize the results of this section. Cosmological gauge-invariant perturbations 
should obey a gauge-invariant action. Using first-order perturbations the action up to third order is the same in the uniform 
energy-density gauge and the flat gauge only after a redefinition of $\zeta_m$ in the uniform energy-density gauge 
$\zeta_m=\zeta_{mc_1}+f_{m1}$ (\ref{rede1}) (the subscript $1$ indicating the use  of only first-order perturbations)  and takes the form 
\bea
S&=&\hat{S}(\zeta_{m})=\tilde{S}(\zeta_{mc_1})=\!\!\int\!\!\d^4x\frac{a\e}{H}(\e\zeta_1-1)\Big((\partial\zeta_1)^2+(\partial\zeta_2)^2\Big)
\nn\\
&+&\!\!\int\!\!\d^4x\Bigg\{\!a^3\!\e H\Bigg[\lh1+\e\zeta_1\rh\lh\dot{\zeta}_1^2+\dot{\zeta}_2^2\rh
-2\partial^i\lambda\lh\dot{\zeta}_2\partial_i\zeta_2+\dot{\zeta}_1\partial_i\zeta_1\rh\nn\\
&&\qquad\qquad\quad-2(\e+\hpa)\zeta_2\partial^i\lambda\partial_i\zeta_2+4\hpe\zeta_2\partial^i\lambda\partial_i\zeta_1
+\frac{1}{2}\zeta_1\lh\partial^i\partial^j\lambda\partial_i\partial_j\lambda-\lh\partial^2\lambda\rh^2\rh\nn\\
&&\qquad\qquad\quad+2\dot{\zeta}_2\lh\chi\zeta_2+\e\zeta_1\lh(\e+\hpa)\zeta_2+\hpe\zeta_1\rh\rh
+\dot{\zeta}_1\lh-4\hpe\zeta_2+\e\zeta_1^2(3\hpa+2\e)\rh\nn\\
&&\qquad\qquad\quad+\zeta_1^2\lh \lh\sqrt{\frac{2\e}{\kappa}}\frac{W_{211}}{H^2}-2\e\lh\e\hpe+\hpa\hpe+\gx^\perp+3\hpe\rh\rh\zeta_2\right.\nn\\
&&\qquad\ \left.\qquad\qquad\ +\lh\sqrt{\frac{2\e}{\kappa}}\frac{W_{111}}{3H^2}-\e\lh\gx^\parallel+3\hpa-(\hpe)^2-(\hpa)^2\rh\rh\zeta_1
\rh\nn\\
&&\qquad\qquad\quad+\zeta_2^2\lh
\sqrt{\frac{2\e}{\kappa}}\frac{W_{221}}{3H^2}-2\e^2-(\hpa)^2+3(\hpe)^2+\frac{2}{3}\hpe\gx^\perp
-3\e(\hpa-\chi)+2\hpa\chi\right.\nn\\
&&\qquad\qquad\ \ \qquad+
\Bigg(\sqrt{\frac{2\e}{\kappa}}\frac{W_{221}}{H^2}+\e\lh-3(\hpe)^2+(\e+\hpa)^2\rh+3\e(\chi-\e-\hpa)\Bigg)\zeta_1\nn\\
&&\left.\qquad\qquad\ \ \qquad+\sqrt{\frac{2\e}{\kappa}}\frac{W_{222}}{3H^2}\zeta_2\rh\Bigg]\Bigg\}
\label{fin}
\eea
where we have kept the notation $\partial^2\lambda=\e\dot{\zeta}_1-2\e\hpe\zeta_2$ in order to mark clearly the terms 
that vanish outside the horizon, namely the terms proportional to $\lambda$ along with the terms involving second-order space derivatives. 
This is one of our main results. 
We managed to compute the cubic action for adiabatic and isocurvature perturbations in the exact theory, beyond any super-horizon or 
slow-roll approximation. 
Its single-field limit coincides with the action computed by Maldacena in \cite{Maldacena:2002vr} or by Rigopoulos in 
\cite{Rigopoulos:2011eq}. 
Let us examine the implications of this action. 
Forgetting about the redefinition of the perturbations in the uniform energy-density gauge, the form of the action is gauge-invariant. 
One can use it to easily calculate the non-Gaussianity related to the interaction terms as is explained in detail 
in \cite{Seery:2005wm,Weinberg:2005vy}. This is known in the literature as $f_{NL}^{(3)}$, the parameter of non-Gaussianity 
related to the three-point correlation function of three first-order perturbations, which is only non-zero in the case of intrinsic 
non-Gaussianity.

However, taking into account the need for a redefinition of the perturbations in the uniform energy-density gauge, one might worry that 
the action is not actually gauge-invariant. 
The action in the uniform energy-density gauge before the redefinition  
has extra terms that are proportional to the second-order equations that the perturbations obey. 
This means that when calculating the non-Gaussianity in the uniform energy-density gauge, one not only has contributions due to the the 
interaction terms in the cubic action, but also ones due to the redefinition of $\zeta_m$, which contribute as explained in (\ref{cyclic}). 
They are part of what is known in the literature as $f_{NL}^{(4)}$, the parameter of non-Gaussianity related to the 
three-point correlation function of a second-order perturbation (in terms of products of first-order ones) and two first-order perturbations, 
which reduces to products of two-point functions of the first-order perturbations. 

This would mean that the non-Gaussianity calculated in the two gauges would not be the same due to the lack of any redefinition in the flat 
gauge. 
However, if one takes into account only the corrections coming from first-order perturbations, the redefinition associated to 
the second-order 
perturbation is not complete as one can check by comparing the super-horizon version of the adiabatic part of (\ref{rede1}) with (\ref{z1ft}).  
As we showed, the solution of this issue is to include second-order fields since they also contribute to the cubic 
action. 
As one would expect these do not change the action itself, so that 
(\ref{fin}) still holds. 
The effect of the new terms is to redefine the perturbations in both gauges. 
It should be noted that, if one had incorporated all quadratic first-order terms (found by a second-order gauge transformation 
as in appendix \ref{app0}) directly as a correction to the first-order perturbations, one would have found the two contributions 
$S_{3(1)}$ and $S_{3(2)}$ together and hence there would have been no initial discrepancy between the two gauges. 
However, we explained at the end of section \ref{cub} our reasons for proceeding in this way. 
So in any case we finally obtain
\be
S=\hat{S}(\zeta_{mc})=\tilde{S}(\zeta_{mc}),\label{finale}
\ee
where $\zeta_{mc}$ is given in (\ref{finu}).  
Now the two redefinitions as well as the action itself are the same for the two gauges, 
hence the action is truly gauge-invariant and the $f_{NL}^{(4)}$, related to the products of first-order $\zeta_m$ in the redefinitions, 
is the same in the two gauges. 

This exact action allows one to compute $f_{NL}^{(4)}$ without the need for 
the slow-roll approximation at horizon crossing that is essential for both the long-wavelength formalism and the $\delta N$ formalism: 
the long-wavelength 
formalism needs slow-roll at horizon crossing in order to allow for the decaying mode to vanish rapidly, while the $\delta N$ 
formalism requires it in order to ignore the derivatives with respect to the canonical momentum. 
Additionally, up to now only 
the slow-roll \textit{field} action \cite{Seery:2005gb}  (and not the action of the $\zeta_m$ themselves) was known, 
so in order to compute the non-Gaussianity at horizon crossing one had to use the long-wavelength 
or $\delta N$ formalism to transform to $\zeta_m$ 
and hence one was in any case required to  make the assumption of slow-roll, 
even if the exact action for the fields would have been known. It will be interesting to investigate models that do not satisfy 
the conditions for the long-wavelength or $\delta N$ formalism using the action (\ref{fin}).

In order to connect the redefinitions to some previously derived results in the literature we assume the super-horizon and slow-roll approximations. The 
super-horizon approximation is already assumed in (\ref{finu}) and it can be supplemented by the condition $\dot{\zeta}_1=2\hpe\zeta_2$. 
The slow-roll assumption translates into $\dot{\zeta}_2=-\chi\zeta_2$. Then 
the quadratic part of the redefinitions, relevant to $f_{NL}^{(4)}$, takes the form 
\bea
&&\zeta_1=\zeta_{1c}+\frac{\e+\hpa}{2}\zeta_1^2+\hpe\zeta_2\zeta_1-\frac{\e+\hpa-\chi}{2}\zeta_2^2\nn\\
&&\zeta_2=\zeta_{2c}+\frac{\hpe}{2}\zeta_1^2+(\e+\hpa-\chi)\zeta_1\zeta_2+\frac{\hpe}{2}\zeta_2^2.
\eea
The redefinitions in this form were used in \cite{Tzavara:2010ge} to find the second-order source term of the evolution equations for the 
super-horizon perturbations. Their contribution to the super-horizon $f_{NL}^{(4)}$ was calculated in that paper using the 
long-wavelength formalism. In the equal-momenta case it was shown to be
\be
-\frac{6}{5}f_{NL,h.c.}^{(4)}=\frac{\es+\hpas+\hpes\bv_{12}}{1+(\bv_{12})^2},\label{difer}
\ee
where the index $*$ indicates the time when the scale exits the horizon and $\bv_{12}$ is essentially a transfer function 
showing how the isocurvature mode $\zeta_2$ sources the adiabatic mode $\zeta_1$ (see \cite{Tzavara:2010ge} for details, where this 
term is part of what is called $g_{sr}$). 
Directly after horizon crossing or equivalently in the 
single-field limit, when $\bv_{12}=0$, this reduces to the well-known result by Maldacena $-6/5f_{NL}^{(4)}=\es+\hpas$. 

\section{Conclusions}\label{concl}

In this paper we settled some unresolved issues concerning gauge invariance at second order in inflation with 
more than one field. Although the gauge-invariant 
curvature perturbation defined through the energy density has been known for many years, 
the energy density is not the quantity that is used in 
calculations of inflationary non-Gaussianity. These use the scalar fields present during inflation instead of their energy. We found this 
gauge-invariant quantity in terms of the fields and discovered that it contains a non-local term unless slow-roll is assumed.

We have also managed to make contact between gauge transformations and the redefinitions of the curvature and 
isocurvature perturbations occurring in the third-order action. 
Since \cite{Maldacena:2002vr} it has been known that the redefinition of the curvature 
perturbation in the action, introduced to remove terms proportional to the first-order equations of motion,  
corresponds to its gauge transformation. 
However, these terms  
appear at first sight to be absent in the flat gauge which would have had as a consequence the absence of 
quadratic contributions of first-order 
curvature perturbations at horizon crossing in this gauge and hence a gauge dependence of the related horizon-crossing 
non-Gaussianity (using Wick's theorem one can calculate the 
three-point correlation function due to these terms, as we did in \cite{Tzavara:2010ge}).  
We have extended the calculation for both gauges to second order and proved that in both of them the 
contributions are 
the same. The difference is that, in our perturbative approach, in the uniform energy-density gauge a part of these contributions is due to the first-order corrections 
and the other part to the second-order fields while in the flat gauge they are all due to the second-order fields. 

In addition to the adiabatic one, we also found the gauge-invariant isocurvature perturbation defined in terms of 
the scalar fields by studying the relevant fully non-linear 
spatial gradient defined in \cite{Rigopoulos:2005xx}. Usually isocurvature perturbations are studied in terms of 
the pressure of the fields. Following \cite{Rigopoulos:2005xx} we found 
a definition using the fields themselves that demonstrates the orthogonality of this quantity to the curvature perturbation. While rewriting the action, these 
isocurvature perturbations appear naturally in the form we have defined them, thus showing that this quantity is the relevant 
one to use during inflation.

In order to achieve the above we computed in section \ref{cubic} the exact cubic action for the perturbations, 
going beyond the 
slow-roll or super-horizon approximations (in appendix \ref{app3} we also give the tensor part of the action). 
This can prove very useful for future calculations. Up to now one had to impose the slow-roll condition at horizon crossing in 
order to calculate the non-Gaussianity. This was because the only two-field action available was that of the fields given in 
\cite{Seery:2005gb}, thus demanding slow-roll at horizon crossing in order to be able to use the long-wavelength 
formalism or the $\delta N$ formalism to find the curvature perturbation bispectrum. 
The action we provide here can be used directly with the in-in formalism \cite{Weinberg:2005vy} in order to 
calculate the exact non-Gaussianity beyond any restrictions, slow-roll or super-horizon.

\appendix
\label{app}

\section{Gauge transformations}
\label{app0}

From the infinite number of possible gauge-invariant combinations, we choose to work with quantities constructed from 
the energy density and the logarithm of the space dependent scale factor $\alpha$. We will consider a gauge transformation 
$\beta_{(i)}=(T_{(i)},\vec{q}_{(i)})$ 
from the hatted gauge 
to the tilded gauge, where for the moment both gauges are taken to be arbitrary (not yet the flat and uniform energy 
density gauge). Notice though, that the space part of the 
transformation is not relevant outside the horizon, since when introduced in the relations below, it is connected to a second-order 
space derivative \cite{Malik:2003mv}. Within the super-horizon approximation, we find using (\ref{trans})
\bea
&&\tilde{\rho}_{(1)}=\hat{\rho}_{(1)}+\dot{\rho}T_{(1)},\qquad 
\tilde{\rho}_{(2)}=\hat{\rho}_{(2)}+\dot{\rho}T_{(2)}+T_{(1)}\lh
2\dot{\hat{\rho}}_{(1)}+\dot{\rho}\dot{T}_{(1)}+\ddot{\rho}T_{(1)}\rh,\nn\\
&&\tilde{\alpha}_{(1)}=\hat{\alpha}_{(1)}+T_{(1)},\qquad
\tilde{\alpha}_{(2)}=\hat{\alpha}_{(2)}+T_{(2)}+T_{(1)}\lh 2\dot{\hat{\alpha}}_{(1)}+\dot{T}_{(1)}\rh.\label{a1}
\eea
We want to construct a gauge-invariant quantity that reduces to $\alpha_{(i)}$ in the uniform energy-density gauge, 
which we now identify with the tilded gauge so that $\tilde{\rho}_{(i)}=0$. This way we find 
\be
T_{(1)}=-\frac{\hat{\rho}_{(1)}}{\dot{\rho}},\qquad\qquad
T_{(2)}=-\frac{\hat{\rho}_{(2)}}{\dot{\rho}}+\frac{\hat{\rho}_{(1)}\dot{\hat{\rho}}_{(1)}}{\dot{\rho}^2}\label{uetr}
\ee
and obtain
\bea
&&\zeta_{1(1)}\equiv\tilde{\alpha}_{(1)}=\hat{\alpha}_{(1)}-\frac{\hat{\rho}_{(1)}}{\dot{\rho}},\nn\\
&&\frac{1}{2}\zeta_{1(2)}\equiv\frac{1}{2}\tilde{\alpha}_{(2)}
=\frac{1}{2}\hat{\alpha}_{(2)}-\frac{1}{2}\frac{\hat{\rho}_{(2)}}{\dot{\rho}}
+\frac{\dot{\hat{\rho}}_{(1)}\hat{\rho}_{(1)}}{\dot{\rho}^2}
-\frac{\hat{\rho}_{(1)}}{\dot{\rho}}\dot{\hat{\alpha}}_{(1)}
-\frac{1}{2}\frac{\hat{\rho}_{(1)}^2}{\dot{\rho}^2} \frac{\ddot{\rho}}{\dot{\rho}}.
\eea
Notice that the initial hatted gauge is still arbitrary, but if one was to associate it with the flat gauge 
$\hat{\alpha}_{(i)}=0$, then the time shift would become $T_{(1)}=\zeta_{1(1)}$.

Next, we derive the exact gauge-invariant adiabatic and isocurvature perturbations, going beyond the super-horizon approximation. 
We use (\ref{trans}) for the scalar fields and the space part of the metric tensor (\ref{metric}), and find
\bea
&&\tilde{\varphi}^A_{(1)}=\hat{\varphi}^A_{(1)}+\dot{\phi}^AT_{(1)}\label{ftrans}\\
&&\tilde{\alpha}_{(1)}\delta_{ij}+\frac{1}{2}\tilde{\gamma}_{(1)ij}
=\hat{\alpha}_{(1)}\delta_{ij}+\frac{1}{2}\hat{\gamma}_{(1)ij}+T_{(1)}\delta_{ij}+\partial_i\partial_jq_{(1)}
+\frac{1}{2}\lh\partial_iq^{\perp}_{(1)j}+\partial_jq^{\perp}_{(1)i}\rh.\label{atrans}
\eea
Here we followed \cite{Bruni:1996im} and split the component $i$ of the space shift as $q_i=\partial_iq+q^{\perp}_i$, where 
$\partial^iq^{\perp}_i=0$. We choose the uniform energy-density gauge, defined by $e_{1A}\tilde{\varphi}^A_{(1)}=0$ and use (\ref{ftrans}) to 
find the first-order time shift to be $T_{(1)}=-e_{1A}\hat{\varphi}^A_{(1)}/\dphi$. Then the trace and the traceless part of 
(\ref{atrans}) give
\bea
&&\tilde{\alpha}_{(1)}=\hat{\alpha}_{(1)}+T_{(1)}+\frac{1}{3}\partial^2q_{(1)}\label{atrans1}\\
&&\tilde{\gamma}_{(1)ij}=\hat{\gamma}_{(1)ij}+2\lh\partial_i\partial_j-\frac{1}{3}\delta_{ij}\partial^2\rh q_{(1)}+
\partial_iq^{\perp}_{(1)j}+\partial_jq^{\perp}_{(1)i}.
\eea
In order to make the definition of the super-horizon adiabatic perturbation at first order (\ref{a1}) to agree with (\ref{atrans1}), 
we choose $q_{(1)}=0$ (any choice of $q_{(1)}$ is a gauge-invariant quantity, but only $q_{(1)}=0$ corresponds with the 
adiabatic perturbation $\zeta_{1(1)}$ in the literature). 
Note that while working with the super-horizon approximation, no such choice needs to be made, and 
$q_{(1)}$ remains arbitrary in that case. Similarly, we also assume that 
$q^{\perp}_{(1)i}=0$, so we find
\be
\zeta_{1(1)}=\tilde{\alpha}_{(1)}=\hat{\alpha}_{(1)}-\frac{1}{\dphi}e_{1A}\hat{\varphi}^A_{(1)}\qquad\mathrm{and}\qquad 
\gamma_{(1)ij}\equiv\tilde{\gamma}_{(1)ij}=\hat{\gamma}_{(1)ij}.
\ee
Using (\ref{ftrans}), one easily finds that the isocurvature perturbation at first order is gauge-invariant since $e_{2A}\dphi^A=0$
\be
\zeta_{2(1)}=-\frac{1}{\dphi}e_{2A}\tilde{\varphi}^A_{(1)}=-\frac{1}{\dphi}e_{2A}\hat{\varphi}^A_{(1)}.
\ee 
We now fix the hatted gauge to be the flat one, $\hat{\alpha}_{(i)}=0$, in order to lighten the calculations.  
This implies that $T_{(1)}=\zeta_{1(1)}$. 

At second order we find
\bea
\tilde{\varphi}^A_{(2)}=\hat{\varphi}^A_{(2)}+T_{(2)}\dphi^A+\zeta_{1(1)}\lh\dot{\zeta}_{1(1)}\dphi^A+\zeta_{1(1)}\ddot{\phi}^A
+2\dot{\hat{\varphi}}_{(1)}^A\rh,\label{ftrans1}
\eea
where we have omitted a term proportional to $q_{(1)}$, which as mentioned above is chosen to be zero. At second order we choose the gauge 
$\frac{1}{2}\tilde{Q}_{1(2)}=
\frac{\e+\hpa}{2}\zeta_{2(1)}^2+\partial^{-2}\partial^i\lh\dot{\zeta}_{2(1)}\partial_i\zeta_{2(1)}\rh$, see (\ref{con}), that reduces to the 
uniform energy-density gauge on super-horizon scales (for the definition of $Q_{1(2)}$ in (\ref{b12}) and details about that 
gauge choice, the reader can refer to appendix \ref{app1}). Using (\ref{difzflat}) we find from (\ref{ftrans1})
\be
T_{(2)}=\hat{Q}_{1(2)}+(\e+\hpa)\lh\zeta_{1(1)}^2-\zeta_{2(1)}^2\rh+\dot{\zeta}_{1(1)}\zeta_{1(1)}-2\hpe\zeta_{1(1)}\zeta_{2(1)}
-2\partial^{-2}\partial^i\lh\dot{\zeta}_{2(1)}\partial_i\zeta_{2(1)}\rh.\label{deft2}
\ee 

Before turning to the adiabatic perturbation, let us prove that the first line of (\ref{z2f}) is a gauge-invariant quantity 
corresponding to the one in (\ref{z2u}). This is true in the exact theory, beyond the long-wavelength approximation, 
for $q_{(1)}=0$. Multiplying (\ref{ftrans1}) with $-e_{2A}/(2\dphi)$, noticing that $e_{2A}\dphi^A=0$, and using 
(\ref{difzflat}), one finds 
\be
\frac{1}{2}\tilde{Q}_{2(2)}=\frac{1}{2}\hat{Q}_{2(2)}+\frac{\hpe}{2}\zeta_{1(1)}^2+\dot{\zeta}_{2(1)}\zeta_{1(1)}
+(\e+\hpa)\zeta_{1(1)}\zeta_{2(1)}\label{x2}
\ee
and indeed by comparing the total gradient of (\ref{z2u}) and (\ref{z2f}) we see that it corresponds to the second-order gauge-invariant 
isocurvature perturbation. 

For the second-order adiabatic perturbation we need to perform the gauge transformation (\ref{trans}) of the space part of the metric tensor 
between the uniform energy-density gauge and the flat gauge
\bea
&&\tilde{\alpha}_{(2)}\delta_{ij}+\frac{1}{2}\tilde{\gamma}_{(2)ij}=\frac{1}{2}\hat{\gamma}_{(2)ij}+T_{(2)}\delta_{ij}+\partial_i\partial_j
q_{(2)}+\frac{1}{2}\lh\partial_iq^{\perp}_{(2)j}+\partial_jq^{\perp}_{(2)i}\rh+\zeta_{1(1)}\dot{\gamma}_{(1)ij}\nn\\
&&\qquad\qquad\qquad\qquad
+\partial_i\zeta_{1(1)}\partial_j\lambda+\partial_j\zeta_{1(1)}\partial_i\lambda+\zeta_{1(1)}\dot{\zeta}_{1(1)}\delta_{ij}
-\frac{1}{a^2H^2}\partial_i\zeta_{1(1)}\partial_j\zeta_{1(1)},\label{atrans2}
\eea
where we substituted $T_{(1)}=\zeta_{1(1)}$ and $T_{(2)}$ is given in (\ref{deft2}). In order to find 
$\zeta_{1(2)}$ we subtract the trace and the $\partial^{-2}\partial^i\partial^j$ of (\ref{atrans2}) to eliminate the term proportional to 
$q_{(2)}$ and obtain the result
\bea
&&\frac{1}{2}\zeta_{1(2)}=\frac{1}{2}\tilde{\alpha}_{(2)}\nn\\
&&\qquad\quad\!=\frac{1}{2}\hat{Q}_{1(2)}+\dot{\zeta}_{1(1)}\zeta_{1(1)}+\frac{\e+\hpa}{2}\lh\zeta_{1(1)}^2-\zeta_{2(1)}^2\rh
-\hpe\zeta_{1(1)}\zeta_{2(1)}
-\partial^{-2}\partial^i\lh\dot{\zeta}_{2(1)}\partial_i\zeta_{2(1)}\rh
\nn\\
&&\qquad\qquad
-\frac{1}{4}\partial^{-2}(\dot{\gamma}_{ij}\partial^i\partial^j\zeta_1)
-\frac{1}{4a^2H^2}\Big((\partial\zeta_1)^2-\partial^{-2}\partial^i\partial^j(\partial_i\zeta_1\partial_j\zeta_1)\Big)\nn\\
&&\qquad\qquad+\frac{1}{2}\Big(\partial^i\lambda\partial_i\zeta_1-\partial^{-2}\partial^i\partial^j(\partial_i\lambda\partial_j\zeta_1)\Big).
\label{adi}
\eea
This is the second-order adiabatic gauge-invariant perturbation in the exact theory, the generalization of (\ref{z1ft}).

\section{Super-horizon calculations}
\label{app1}

In the first part of this appendix we give the detailed calculations of subsection \ref{gaugetrans}, while in the second part 
we show the ones relevant to subsection \ref{grad}.

An important property of the long-wavelength assumption is that outside the horizon the uniform energy density can be 
recast in terms of the fields at least at first order: one can show 
that the exact 0i-Einstein equation (\ref{fieldeqsu}) outside the horizon can be rewritten as 
\be
\partial_i\bar{\rho}=-3\bH\bar{\Pi}_B\partial_i\varphi^B.\label{0i}
\ee
Again $\bar{\rho}$ denotes the fully non-linear energy density $\bar{\rho}=\bar{\Pi}^2/2+W$. 
Expanding (\ref{0i}) to first 
order and using the background equations to prove that
\be
\dot{\rho}=-3\Pi^2NH,
\ee
one can show that outside the horizon
\be
\frac{\rho_{(1)}}{\dot{\rho}}=\frac{1}{\Pi N}e_{1A}\varphi^A_{(1)}
\label{ratio1}
\ee
so that 
\be
\zeta_{1(1)}=\tilde{\alpha}_{(1)}=\alpha_{(1)}-\frac{H}{\Pi}e_{1A}\varphi^A_{(1)}.
\ee
and thus the energy-density constraint $\tilde{\rho}_{(1)}=0$ is equivalent to $e_{1A}\tvarphi^A_{(1)}=0$.
 
Unfortunately the nice property described by (\ref{ratio1}) does not hold anymore at second order. After expanding up to 
second order, combining the second-order Einstein equations and using the completeness relation of the field basis, 
one can show that (note that the zeroth order lapse function is taken from now on to be $N(t)=1/H(t)$) 
\be
\frac{\rho_{(2)}}{\dot{\rho}}-\frac{1}{\dphi}e_{1A}\varphi^A_{(2)}
=\lh\frac{1}{\dphi}e_{1A}\varphi_{(1)}^A\rh^2(\hpa-\e)+\frac{1}{\dphi^2}A,
\label{ratio2}
\ee
with 
\bea
\frac{1}{2}\partial_iA&=&\hpe e_{2B}\varphi^B_{(1)}e_{1A}\partial_i\varphi^A_{(1)}+e_{2B}\partial_i\varphi^B_{(1)}\e_{2A}\dvphi^A_{(1)},
\label{alp}
\eea
where we used (\ref{dere}) to simplify the expressions in terms of the slow-roll parameters.
We see that the purely second-order contribution of $\rho_{(2)}$ is recast as a second-order contribution of $\varphi^A_{(2)}$, 
some quadratic first-order terms and a non-local term arising essentially from the 0i-Einstein equation.
In the flat gauge $\partial_i\ha=0$ this non-local term can be written as
\be
\frac{1}{2\phi^2}\partial_i\hat{A}=\partial_i\Bigg[\hpe\zeta_{1(1)}\zeta_{2(1)}+\frac{\e+\hpa}{2}\zeta_{2(1)}^2\Bigg]
+\dot{\zeta}_{2(1)}\partial_i\zeta_{2(1)},\label{alp_flat}
\ee
where when needed we employed the following useful relations (valid for the flat gauge beyond the long-wavelength approximation):
\bea
-\frac{1}{\dphi}e_{1A}\dot{\hat{\varphi}}^A_{(1)}&=&\dot{\zeta}_{1(1)}+(\e+\hpa)\zeta_{1(1)}-\hpe\zeta_{2(2)},\nn\\
-\frac{1}{\dphi}e_{2A}\dot{\hat{\varphi}}^A_{(1)}&=&\dot{\zeta}_{2(1)}+(\e+\hpa)\zeta_{2(1)}+\hpe\zeta_{1(2)},\label{difzflat}
\eea
derived by differentiating the adiabatic perturbation and the  
new combination
\be
\zeta_{2(1)}\equiv-\frac{H}{\Pi}e_{2A}\varphi^A_{(1)},
\ee
that represents the isocurvature perturbation to first order.
Putting everything together in (\ref{36}) we find the second-order gauge-invariant curvature 
perturbation in the flat gauge to be
\bea
\frac{1}{2}\zeta_{1(2)}=\frac{1}{2}\tilde{\alpha}_{(2)}&=&\frac{1}{2}\hat{Q}_{1(2)}
+\frac{\e+\hpa}{2}\zeta_{1(1)}^2-\frac{\e+\hpa}{2}\zeta_{2(1)}^2-\hpe\zeta_{1(1)}\zeta_{2(1)}+\dot{\zeta}_{1(1)}\zeta_{1(1)}\nn\\
&-&\partial^{-2}\partial^i\lh\dot{\zeta}_{2(1)}\partial_i\zeta_{2(1)}\rh,
\eea
where we defined the auxiliary quantities
\be
Q_{m(i)}\equiv-\frac{H}{\Pi}e_{mA}\varphi^A_{(i)}.\label{b12}
\ee

We turn now to the calculations relevant to the gradient of the perturbations. 
In the uniform energy-density gauge we can use (\ref{0i}) to find the constraints
\bea
&&\partial_i\tilde{\rho}_{(1)}=-3\dphi_A\partial_i\tvarphi^A_{(1)}=0 \qquad \mathrm{or} \qquad \partial_i\tilde{Q}_{1(1)}=0\label{con1}\\
&&\partial_i\tilde{\rho}_{(2)}=-\frac{3}{2}\dphi_A\partial_i\tvarphi^A_{(2)}-3\tdvphi_{A(1)}\partial_i\tvarphi^A_{(1)}=0\qquad \mathrm{or}\nn\\
&&\frac{1}{2}\partial_i\tilde{Q}_{1(2)}=\frac{1}{\dphi^2}\tdvphi_{A(1)}\partial_i\tvarphi^A_{(1)}=
\frac{\e+\hpa}{2}\partial_i\zeta_{2(1)}^2+\dot{\zeta}_{2(1)}\partial_i\zeta_{2(1)},
\label{con}
\eea
so that when inspecting (\ref{akoma}) we see that
\begin{displaymath}
\frac{1}{2}\tilde{\zeta}_{1i(2)}=\frac{1}{2}\partial_i\tilde{\alpha}_{(2)}.
\end{displaymath}
To derive the last equality in (\ref{con}) we used the completeness relation of the field basis along with the following relations 
valid beyond the long-wavelength approximation:
\bea
&&-\frac{1}{\dphi}e_{1A}\dot{\tilde{\varphi}}^A_{(1)}=-\hpe\zeta_{2(1)},\nn\\
&&-\frac{1}{\dphi}e_{2A}\dot{\tilde{\varphi}}^A_{(1)}=\dot{\zeta}_{2(1)}+(\e+\hpa)\zeta_{2(1)},\label{difzue}
\eea
derived by differentiating the definition of the isocurvature perturbation and the first-order uniform energy-density gauge constraint 
$e_{1A}\tilde{\varphi}^A_{(1)}=0$.

\section{Second-order action calculation}
\label{app2}

In order to rewrite the action we first need to calculate the 
extrinsic curvature. To do that we decompose $\bar{N}=1/H+N_1$, $N^i=\partial^i\psi+N^i_{\perp}$, 
where $\partial_iN^i_{\perp}=0$. 
 From now on we drop the explicit 
subscript $(1)$ on first-order quantities and set $\kappa^2=1$ to lighten the notation 
(notice though that the final results remain unchanged when we restore $\kappa^2=8\pi G$, 
since all $\kappa^2$ are absorbed in $\e$ when rewriting the fields in terms of $\zeta_m$). We start by performing the calculation in the gauge 
\be
e_{1A}\tilde{\varphi}^A=0,\label{ugauge}
\ee
which we call uniform energy-density gauge, 
since the above constraint reduces to zero energy perturbation outside the horizon.
We first use the energy and momentum constraint (\ref{energy}), (\ref{momentum}) to find that to first order
\bea
&&\tilde{N}_1=\frac{\dot{\tilde{\alpha}}}{H}=\frac{\dot{\zeta}_1}{H}, \quad\ \ \qquad \tilde{N}^i_{\perp}=0,\nn\\
&&\tilde{\psi}=-\frac{1}{a^2}\frac{\zeta_1}{H^2}+\lambda, \qquad \partial^2\lambda=\e\dot{\zeta}_1-2\e\hpe\zeta_2.
\label{n1ni}
\eea
It turns out that we do not need to calculate the shift or the lapse function to higher order, since in the action 
those terms are multiplied by constraint relations and hence vanish.

We start by working out the scalar part of the action. 
Keeping in mind the gauge constraint (\ref{ugauge}) we perturb (\ref{actionexact}) to second order
\bea
\tilde{S}_2=\frac{1}{2}\int\d^4x\Bigg\{&&\!\!\!\!\!a^3 e^{3\tilde{\alpha}}
\Big[(\frac{1}{H}+\tilde{N}_1+\frac{\tilde{N}_2}{2})\Big(-2W-2W_A\tvarphi^A-W_{AB}\tvarphi^A\tvarphi^B\Big)\nn\\
&&\quad\quad+H(1-H\tilde{N}_1+H^2\tilde{N}_1^2-H\frac{\tilde{N}_2}{2})
\Big(-6(1+\dot{\tilde{\alpha}})^2+\dot{\phi}^2+2\dot{\phi}^A\dot{\tvarphi}_A+\dot{\tvarphi}^2\Big)\Big]\nn\\
&&\!\!\!\!\!-a e^{\tilde\alpha}\Big[(\frac{1}{H}+\tilde{N}_1)2
\lh(\partial\tilde{\alpha})^2+2\partial^2\tilde{\alpha}\rh
+\frac{1}{H}\partial_i\tilde{\varphi}_A\partial^i\tilde{\varphi}^A\Big]\Bigg\},
\eea
where we have omitted a total derivative with respect to $\tilde{\psi}$. 
We then use the background Einstein and field equations to eliminate some terms and find that the term 
proportional to $\tilde{N}_2$ vanishes.
Now the second-order action can be written as
\bea
\tilde{S}_2=\frac{1}{2}\!\!\int\!\!\d^4x\Bigg\{\!\!\!\!&& a^3H\Big[
\dot{\zeta}_1\Big(\!-4\dot{\phi}^A\dot{\tvarphi}_A
+\dot{\phi}^2\dot{\zeta}_1\Big)+9\zeta^2_1(-6+\dot{\phi}^2)-\frac{1}{H^2}W_{AB}\tvarphi^A\tvarphi^B-36\zeta_1\dot{\zeta}_1
+\dot{\tvarphi}^2\Big]\nn\\
&&\!\!\!\!-a\frac{1}{H}\Big[2\e(\partial\zeta_1)^2+\partial^i\tilde{\varphi}_A\partial_i\tilde{\varphi}^A\Big]\Bigg\}.
\label{sin}
\eea
The terms of (\ref{sin}) proportional to $\tilde{\varphi}^A$ can be recast in terms of the curvature perturbations by applying the 
completeness property of the field basis and (\ref{difzue}),
so that after integrating by parts and using $\dot{H}=-\e H$ it can be written as
\bea
\tilde{S}_2=\int\d^4x\ \e\Bigg\{&&\!\!\!\!a^3H\Big[\dot{\zeta}_1^2+\dot{\zeta}_2^2-4\hpe\dot{\zeta}_1\zeta_2
-3(\chi-\e-\hpa)\zeta_2^2
+2(\e+\hpa)\zeta_2\dot{\zeta}_2\nn\\
&&\qquad+(\hpe)^2\zeta_2^2+(\e+\hpa)^2\zeta_2^2\Big]-
a\frac{1}{H}\Big[ (\partial\zeta_1)^2+(\partial\zeta_2)^2\Big]
\Bigg\}\label{simple}
\eea
or after further integration by parts
\bea
\tilde{S}_2\!=\!\!\int\!\!\d^4x\ \e \Bigg\{&&\!\!\!\!\!-a\frac{1}{H}\Big( (\partial\zeta_1)^2+(\partial\zeta_2)^2\Big)
 +a^3H\Big(\dot{\zeta}_1^2+\dot{\zeta}_2^2-4\hpe \dot{\zeta}_1\zeta_2
+2\chi \dot{\zeta}_2\zeta_2\Big)\nn\\
&&\!\!\!\!\!\!\!\!\!+a^3H\Big(\!\sqrt{\frac{2\e}{\kappa}}\frac{W_{221}}{3H^2}\!-\!2\e^2\!-\!(\hpa)^2\!+\!3(\hpe)^2\!+\!\frac{2}{3}\hpe\gx^\perp
\!-\!3\e(\hpa\!-\!\chi)\!+\!2\hpa\chi\Big) \zeta_2^2\Bigg\}.\label{s22}
\eea

We can reach the same result while working in the flat gauge $\partial_i{\hat{\alpha}}=0$.  
One can prove that $\hat{N}_1=-\e\zeta_1/H,\ \hat{N}^i_{\perp}=0$ and 
$\partial^2\hat{\psi}=\partial^2\lambda=\e\dot{\zeta}_1-2\e\hpe\zeta_2$. The $\hat{\psi}$ terms cancel out and the second-order 
action takes the form
\bea
 \hat{S}_2&=&\frac{1}{2}\int\d^4x\Big\{ a^3\Big[H\dot{\hat{\varphi}}^2-\frac{1}{H}W_{AB}\hat{\varphi}^A\hat{\varphi}^B
+\hat{N}_1(-2W_A\hat{\varphi}^A-2H^2\dphi^A\dot{\hat{\varphi}}_A)+\hat{N}_1^2H^3(-6+\dphi^2)\Big]\nn\\
&&\qquad\qquad\!\!-a\frac{1}{H}\partial^i\hat{\varphi}_A\partial_i\hat{\varphi}^A\Big\}.
\eea
Using the definition of $\zeta_m$, along with the background equations (\ref{fieldeq}), (\ref{srvar}) 
and (\ref{difzflat}) this can be rewritten as (\ref{s22}). 

The second-order tensor part of the action in both gauges takes the form 
\be
S_{2\gamma}=\int\d^4xL_{2\gamma}=\frac{1}{2}\int\d^4x\Big\{\frac{a^3}{4}H(\dot{\gamma}_{ij})^2
-\frac{a}{4H}(\partial_k\gamma_{ij})^2\Big\},
\ee 
where $L_{2\gamma}$ is the second-order Lagrangian for the tensor modes. 
We also give the equation of motion of the gravitational waves
\be
\frac{\delta  L_{2\gamma}}{\delta\gamma_{ij}}=-\frac{1}{4}\frac{\d}{\d t}(a^3H\dot{\gamma}_{ij})
+\frac{1}{4}\frac{a}{H}\partial^2\gamma_{ij}=0,
\ee
which we are going to use in the next section. In this paper we will not discuss the evolution and physics of gravitational 
waves, but at linear order this is 
a standard subject in 
the literature, for a discussion see for example \cite{Misner:1974qy}.

\section{Third-order action calculation}
\label{app3}

In order to compute $S_3$ we follow the same procedure starting from the uniform energy-density gauge. 
Notice that $\tilde{N}_3$ will multiply $(-2W+6H^2-\Pi^2)$ in exact 
analogy with $\tilde{N}_2$ in $S_2$, so it vanishes. Moreover, the overall factor multiplying $\tilde{N}_2$ is 
the first-order energy constraint (\ref{n1ni}), so it can be consistently set to zero as well.

We start by computing the cubic action of the first-order curvature perturbations up to $\tilde{N}_1$ involving only 
scalar quantities
\bea
\tilde{S}_{3(1)}\!=\!\frac{1}{2}\int\!\!\d^4x&&\!\!\!\!\Bigg\{\!a^3e^{3\tilde{\alpha}}\Bigg[ (\frac{1}{H}+\tilde{N}_1)
           \Big(-2W-2W_A\tvarphi^A-W_{AB}\tvarphi^A\tvarphi^B-\frac{1}{3}W_{ABC}\tvarphi^A\tvarphi^B\tvarphi^C\Big)\nn\\
&&\qquad+H\Big[(1-H\tilde{N}_1+H^2\tilde{N}_1^2
-H^3\tilde{N}_1^3
)\Big(-6(1+\dot{\tilde{\alpha}})^2
+\dot{\phi}^2+2\dot{\phi}^A\dot{\tvarphi}_A+\dot{\tvarphi}^2\Big)\nn\\
&&\qquad+\Big(\partial^i\partial^j\tilde{\psi}\partial_i\partial_j\tilde{\psi}
-(\partial^2\tilde{\psi})^2\Big)(1-H\tilde{N}_1)-4\partial^i\tilde{\psi}\partial_i\zeta_1\partial^2\tilde{\psi}
-2\dot{\tilde{\varphi}}_A\partial^i\tilde{\psi}\partial_i\tilde{\varphi}^A\Big]\Bigg]\nn\\
&&\!\!\!\!\!-ae^{\tilde{\alpha}}\Big[(\frac{1}{H}+\tilde{N}_1)\Big(\partial^i\tilde{\varphi}_A\partial_i\tilde{\varphi}^A
+4\partial^2\zeta_1+2(\partial\zeta_1)^2\Big)
  \Big]\Bigg\}.\label{s311}
\eea
After using the background equations and the definitions of the perturbations, eq. (\ref{s311}) takes the form
\bea
\tilde{S}_{3(1)}=\!\!\int\!\!\d^4x\Bigg\{&&\!\!\!\!\!\!a^3e^{3\zeta}H\Big[\e(1-\dot{\zeta}_1)
\Big(\!\dot{\zeta}_1^2+\dot{\zeta}_2^2+2(\e+\hpa)\dot{\zeta}_2\zeta_2+\!\lh\!(\hpe)^2\!+\!(\e+\hpa)^2\rh\zeta_2^2
-2\hpe \zeta_2\dot{\zeta}_1\!\!\Big)\nn\\
&&\quad-2\e\hpe \zeta_2\dot{\zeta}_1
-3(1+\dot{\zeta}_1)\e(\chi-\e-\hpa)\zeta_2^2+\e\sqrt{\frac{2\e}{\kappa}}\frac{W_{222}}{3H^2}\zeta_2^3
-2\partial^i\tilde{\psi}\partial_i\zeta_1\partial^2\tilde{\psi}\nn\\
&&\quad+\frac{1}{2}\Big(\partial^i\partial^j\tilde{\psi}\partial_i\partial_j\tilde{\psi}
-(\partial^2\tilde{\psi})^2\Big)(1-\dot{\zeta}_1)
-2\e\partial^i\tilde{\psi}\Big((\e+\hpa)\zeta_2\partial_i\zeta_2+\dot{\zeta}_2\partial_i\zeta_2\Big)\Big]\nn\\
&&\!\!\!\!\!\!\!\!\!\!\!
-a\frac{1}{H}(\zeta_1+\dot{\zeta}_1)\Big[2\partial^2\zeta+(\partial\zeta_1)^2+\e(\partial\zeta_2)^2\Big]\Bigg\}.\label{s31}
\eea
By performing integrations by parts in (\ref{s31}) we find 
\bea
\tilde{S}_{3(1)}\!=\!\!\int\!\!\d^4x\!\!&\Bigg\{&\!\!\!\!a^3\e H\Bigg[\e\zeta_1(\dot{\zeta}_1^2+\dot{\zeta}_2^2)
-2\dot{\zeta}_1\partial^i\lambda\partial_i\zeta_1-2\dot{\zeta}_2\partial^i\lambda\partial_i\zeta_2
-2\Big(\e\hpa\!+\!(\hpa)^2\!+\!(\hpe)^2\!\Big)\zeta_1\zeta_2\dot{\zeta}_2\nn\\
&&
\quad\ \ +(3\e\hpe+\gx^\perp)\zeta_1^2\dot{\zeta}_2
+2(\e\hpe+\gx^\perp)\zeta_1\zeta_2\dot{\zeta}_1
-\Big(\e^2+2\e\hpa+(\hpa)^2+(\hpe)^2\Big)\zeta_2^2\dot{\zeta}_1 \nn\\
&&
\quad\ \ 
\nn\\
&&\quad\ \ +\Big(2\e^2+3\e\hpa-\!(\hpa)^2-\!(\hpe)^2+\gx^\parallel\Big)\zeta_1^2\dot{\zeta}_1
+\sqrt{\frac{2\e}{\kappa}}\frac{W_{222}}{3H^2}\zeta_2^3 \nn\\
&&\quad\ \ +\Bigg(2(\e+\hpa)\gx^\perp+\hpe\Big(2\e(3+\e)+6\hpa-\gx^\parallel-3\chi\Big)\Bigg)\zeta_1^2\zeta_2
\nn\\
&&\quad\ \ +\Bigg(\!\!-\e\Big(4\e^2+6\e+12\e\hpa+\hpa(9+8\hpa)+8(\hpe)^2+2\gx^\parallel-3\chi\Big)
\!+\!\sqrt{\frac{2\e}{\kappa}}\frac{W_{221}}{H^2}\nn\\
&&\qquad\ \ -3(\hpa)^2-3(\hpe)^2-2\hpa\gx^\parallel-2\hpe\gx^\perp\Bigg)\zeta_2^2\zeta_1
-2(\e+\hpa)\zeta_2\partial^i\lambda\partial_i\zeta_2\nn\\
&&\quad\ \ +4\hpe\zeta_2\partial^i\lambda\partial_i\zeta_1
+\frac{1}{2}\zeta_1(\partial^i\partial^j\lambda\partial_i\partial_j\lambda-(\partial^2\lambda)^2)\Bigg]
+a\e^2\frac{1}{H}\zeta_1\Big[(\partial\zeta_1)^2\!+\!(\partial\zeta_2)^2
\Big]\nn\\
&&\!\!\!\!\!\!\!\!\!\!-\frac{\delta L_2}{\delta\zeta_1}\Big(\frac{\e+\hpa}{2}\zeta_1^2-\hpe\zeta_1\zeta_2
+\dot{\zeta}_1\zeta_1-\frac{1}{4a^2H^2}(\partial\zeta_1)^2
+\frac{1}{4a^2H^2}\partial^{-2}\partial^i\partial^j(\partial_i\zeta_1\partial_j\zeta_1)\nn\\
&&\quad+\frac{1}{2}\partial^i\zeta_1\partial_i\lambda
-\frac{1}{2}\partial^{-2}\partial^i\partial^j(\partial_i\lambda\partial_j\zeta_1)\Big)\!-\!\frac{\delta L_2}{\delta\zeta_2}
\lh\!(\e+\hpa)\zeta_1\zeta_2+\dot{\zeta}_2\zeta_1+\frac{\hpe}{2}\zeta_1^2\rh\!\!\Bigg\},
\eea
where $\delta L_2/\delta\zeta_m$ are the first-order equations of motion. 
We can further integrate by parts the rest of the action to simplify it and prove that it takes the form of the 
flat gauge action (\ref{sflatap}), as expected since the action should be gauge-invariant.
The terms involving $\lambda$ along with the terms with 
space gradients vanish outside the horizon in the long-wavelength approximation, since $\lambda$ is equal to the 
first-order super-horizon energy constraint (\ref{mc}).

Finally we include the second-order fields. The extra terms in the action are
\bea
\tilde{S}_{3(2)}\!=\!\frac{1}{2}\int\!\!\d^4x\!&\Bigg\{&\!\!\!\!a^3e^{3\zeta_1}\!\Big[(\frac{1}{H}+\tilde{N}_1)
(-W_A\tvarphi^A_{(2)}\!-\!W_{AB}
\tvarphi^A_{(2)}\tvarphi^B)\!+\!H(1-H\tilde{N}_1)(\dphi_A\tdvphi^A_{(2)}+\tdvphi_A\tdvphi^A_{(2)})\nn\\
&&
\qquad-H\dot{\phi}^A\partial^i\tilde{\psi}\partial_i\tilde{\varphi}_{(2)}+2H\dot{\zeta}_{1(2)}\partial^2\tilde{\psi}\Big]\nn\\
&&
\!\!\!\!\!\!\!\!-a\frac{1}{H}\Big[-\partial^i\zeta_1\partial_i\zeta_{1(2)}+\dot{\zeta}_1\partial^2\zeta_{1(2)}
+\partial^i\tilde{\varphi}^A\partial_i\tilde{\varphi}_{A(2)}\Big]\Bigg\},\label{s32}
\eea
where $\varphi^A$ without a subscript always denotes the first-order perturbation.  
After performing integrations by parts we find
\bea
\tilde{S}_{3(2)}=\int\d^4x\Bigg\{\frac{\delta L_2}{\delta\zeta_1}\lh\frac{\zeta_{1(2)}}{2}+\frac{\tilde{Q}_{1(2)}}{2}\rh
+\frac{\delta L_2}{\delta\zeta_2}\frac{\tilde{Q}_{2(2)}}{2}\Bigg\}.\label{s32s}
\eea

Next, we perform the same calculation for the flat gauge, starting from
\bea
\hat{S}_{3(1)}=\frac{1}{2}\int\d^4x\!\!\!\!&&\Bigg\{a^3\Big[ (\frac{1}{H}+\hat{N}_1)
           \Big(-2W-2W_A\hat{\varphi}^A-W_{AB}\hat{\varphi}^A\hat{\varphi}^B-\frac{1}{3}W_{ABC}\hat{\varphi}^A\hat{\varphi}^B
\hat{\varphi}^C\Big)\\
 &&+H(1-H\hat{N}_1+H^2\hat{N}_1^2
-H^3\hat{N}_1^3
)\Big(-6+\dot{\phi}^2+2\dot{\phi}^A\dot{\hat{\varphi}}_A+\dot{\hat{\varphi}}^2
+4\partial^2\hat{\psi}\nn\\
&&+\partial^i\partial^j\hat{\psi}\partial_i\partial_j\hat{\psi}-(\partial^2\hat{\psi})^2
-2\partial^i\hat{\psi}\dot{\phi}^A\partial_i\hat{\varphi}_A
-2\partial^i\hat{\psi}\dot{\hat{\varphi}}^A\partial_i\hat{\varphi}_A\Big)\Big]
-a\hat{N}_1\partial^i\hat{\varphi}^A\partial_i\hat{\varphi}_A\Bigg\},\nn
\eea
again taking into account that $\hat{N}_2$ multiplies the first-order energy constraint and thus we set it to zero. 
We find using the definition of $\zeta_m$, along with (\ref{fieldeq}), (\ref{srvar}) and (\ref{difzflat})
\bea
\hat{S}_{3(1)}&=&\!\!\int\!\!\d^4x\Bigg\{\!a^3\!\e H\Bigg[\e\zeta_1(\dot{\zeta}_1^2+\dot{\zeta}_2^2)
-2\dot{\zeta}_2\partial^i\lambda\partial_i\zeta_2-2\dot{\zeta}_1\partial^i\lambda\partial_i\zeta_1
\nn\\
&&\qquad\qquad\quad+2\e(\e+\hpa)\zeta_1\zeta_2\dot{\zeta}_2
+2\e\hpe\zeta_1^2\dot{\zeta}_2 
+\e(3\hpa+2\e)\zeta_1^2\dot{\zeta}_1 \nn\\
&&\qquad\qquad\quad+\Bigg(\!\!\sqrt{\frac{2\e}{\kappa}}\frac{W_{211}}{H^2}-2\e(\e\hpe+\hpa\hpe+\gx^\perp+3\hpe)\Bigg)\zeta_1^2\zeta_2
+\sqrt{\frac{2\e}{\kappa}}\frac{W_{222}}{3H^2}\zeta_2^3\nn\\
&&\qquad\qquad\quad+\Bigg(\!\!\sqrt{\frac{2\e}{\kappa}}\frac{W_{221}}{H^2}+\e\lh-3(\hpe)^2+(\e+\hpa)^2\rh+3\e(\chi-\e-\hpa)\Bigg)\zeta_2^2\zeta_1\nn\\
&&\qquad\qquad\quad+\lh\!\!\sqrt{\frac{2\e}{\kappa}}\frac{W_{111}}{3H^2}-\e\lh\gx^\parallel+3\hpa-(\hpe)^2-(\hpa)^2\rh\rh\zeta_1^3
\nn\\
&&\qquad\qquad\quad-2(\e+\hpa)\zeta_2\partial^i\lambda\partial_i\zeta_2+4\hpe\zeta_2\partial^i\lambda\partial_i\zeta_1
+\frac{1}{2}\zeta_1(\partial^i\partial^j\lambda\partial_i\partial_j\lambda-(\partial^2\lambda)^2)\Bigg]
\nn\\
&&\qquad\!+\frac{a\e^2}{H}\zeta_1\Big((\partial\zeta_1)^2+(\partial\zeta_2)^2\Big)\Bigg\}.\label{sflatap}
\eea

Finally we include the second-order fields. The surviving terms in the action are
\bea
\hat{S}_{3(2)}\!=\!\frac{1}{2}\int\!\!\d^4x\!&\Bigg\{&\!\!\! a^3\Big[(\frac{1}{H}+\hat{N}_1)(-W_A\hat{\varphi}^A_{(2)}-W_{AB}
\hat{\varphi}^A_{(2)}\hat{\varphi}^B)
+H(1-H\hat{N}_1)(\dphi_A\dot{\hat{\varphi}}^A_{(2)}+\dot{\hat{\varphi}}_A\dot{\hat{\varphi}}^A_{(2)})\nn\\
&&\ \ -H\dot{\phi}^A\partial^i\hat{\psi}\partial_i\hat{\varphi}_{A(2)}\Big]
-a\frac{1}{H}\partial^i\hat{\varphi}^A\partial_i\hat{\varphi}_{A(2)}\Bigg\}
\eea
and they can be rewritten as 
\bea
\hat{S}_{3(2)}=\int\d^4x\Bigg\{\frac{\delta L_2}{\delta\zeta_1}\frac{\hat{Q}_{1(2)}}{2}
+\frac{\delta L_2}{\delta\zeta_2}\frac{\hat{Q}_{2(2)}}{2}\Bigg\}.
\eea

In the last part of this appendix we consider the tensor scalar part of the action. There will be no contributions from the 
second-order fields, since these 
cancel due to $\gamma_{ij}$ being transverse. We start from the action for two scalar and one tensor modes in the 
uniform energy-density gauge
\bea
\tilde{S}_{\zeta\zeta\gamma}=\int\d^4x&\Big\{&\frac{a}{H}\Big[-2\gamma_{ij}\partial^i\dot{\zeta}_1\partial^j\zeta_1
-\gamma_{ij}\partial^i\zeta_1\partial^j\zeta_1+\e\gamma_{ij}\partial^i\zeta_2\partial^j\zeta_2\Big]\nn\\
&&\!\!\!\!\!\!\!+\frac{1}{2}a^3H\Big[-(3\zeta_1-\dot{\zeta}_1)\dot{\gamma}_{ij}\partial^i\partial^j\tilde{\psi}
+\partial_k\gamma_{ij}\partial^i\partial^j\tilde{\psi}\partial^k\tilde{\psi}\Big]\Big\},
\eea
which after integrations by parts becomes
\bea
\tilde{S}_{\zeta\zeta\gamma}\!\!=\!\!\!\int\!\!\d^4x\!&\Big\{&\!\!\!a^3H\Big[
\frac{\e}{2}\dot{\gamma}_{ij}\partial^i\zeta_1\partial^j\lambda
+\frac{1}{4}\partial^2\gamma_{ij}\partial^i\lambda\partial^j\lambda\Big]
+\frac{a}{H}\e\gamma_{ij}\Big[\partial^i\zeta_1\partial^j\zeta_1+
\partial^i\zeta_2\partial^j\zeta_2\Big]\label{szzg}\\
&&\!\!\!\!\!\!+\frac{\delta L_{2\gamma}}{\delta\gamma_{ij}}\Big(\frac{1}{a^2H^2}\partial_i\zeta_1\partial_j\zeta_1
-(\partial_i\zeta_1\partial_j\lambda+\partial_j\zeta_1\partial_i\lambda)\Big)
+\frac{\delta L_2}{\delta\zeta_1}\frac{1}{4}\partial^{-2}(\dot{\gamma}_{ij}\partial^i\partial^j\zeta_1)\Big\}.\nn
\eea
In the flat gauge one can find directly after substitution in (\ref{actionexact}) the first line of (\ref{szzg}), so that 
there are no redefinitions.

Finally we calculate the part of the action consisting of one scalar and two tensor modes, starting from the uniform 
energy-density gauge
\bea
\tilde{S}_{\zeta\gamma\gamma}=\frac{1}{2}\int\d^4x\Big\{a^3H\Big[\frac{1}{4}(3\zeta_1-\dot{\zeta}_1)
(\dot{\gamma}_{ij})^2-\frac{1}{2}\dot{\gamma}_{ij}
\partial_k\gamma^{ij}\partial^k\tilde{\psi}\Big]-\frac{a}{4H}(\zeta_1+\dot{\zeta}_1)(\partial_k\gamma_{ij})^2\Big\}
\eea
or equivalently
\bea
\tilde{S}_{\zeta\gamma\gamma}&=&\int\d^4x\Big\{-\zeta_1\dot{\gamma}_{ij}\frac{\delta L_{2\gamma}}{\delta\gamma_{ij}}
+a^3H\Big[\frac{\e}{8}\zeta_1 (\dot{\gamma}_{ij})^2-\frac{1}{4}\dot{\gamma}_{ij}\partial_k\gamma^{ij}\partial^k\lambda\Big]
+\frac{a}{8H}\e\zeta_1(\partial_k\gamma_{ij})^2\Big\},
\eea
while in the flat gauge we find directly
\bea
\hat{S}_{\zeta\gamma\gamma}=\frac{1}{2}\int\d^4x\Big\{a^3H\Big[\frac{\e}{4}\zeta_1(\dot{\gamma}_{ij})^2-\frac{1}{2}\dot{\gamma}_{ij}
\partial_k\gamma^{ij}\partial^k\lambda\Big]+\frac{a}{4H}\e\zeta_1(\partial_k\gamma_{ij})^2\Big\}.
\eea

The three tensor modes action does not contain any redefinitions. For details the reader may look in \cite{Maldacena:2002vr}.

\bibliography{bib}{}

\providecommand{\href}[2]{#2}\begingroup\raggedright\begin{thebibliography}{10}

\bibitem{paper1}
A.~H. Guth and S.~Y. Pi, ``{Fluctuations in the New Inflationary Universe}'',
\href{http://dx.doi.org/10.1103/PhysRevLett.49.1110}{{\em Phys. Rev. Lett.}
  {\bf 49} (1982)  1110--1113}.

\bibitem{paper2}
S.~W. Hawking, ``{The Development of Irregularities in a Single Bubble
  Inflationary Universe}'',
\href{http://dx.doi.org/10.1016/0370-2693(82)90373-2}{{\em Phys. Lett.} {\bf
  B115} (1982)  295}.

\bibitem{paper3}
A.~A. Starobinsky, ``{Dynamics of Phase Transition in the New Inflationary
  Universe Scenario and Generation of Perturbations}'',
\href{http://dx.doi.org/10.1016/0370-2693(82)90541-X}{{\em Phys. Lett.} {\bf
  B117} (1982)  175--178}.

\bibitem{paper4}
J.~M. Bardeen, P.~J. Steinhardt, and M.~S. Turner, ``{Spontaneous Creation of
  Almost Scale - Free Density Perturbations in an Inflationary Universe}'',
\href{http://dx.doi.org/10.1103/PhysRevD.28.679}{{\em Phys. Rev.} {\bf D28}
  (1983)  679}.

\bibitem{Komatsu:2010fb}
{\bf WMAP} Collaboration, E.~Komatsu {\em et al.}, ``{Seven-Year Wilkinson
  Microwave Anisotropy Probe (WMAP) Observations: Cosmological
  Interpretation}'', \href{http://dx.doi.org/10.1088/0067-0049/192/2/18}{{\em
  Astrophys.J.Suppl.} {\bf 192} (2011)  18},
  \href{http://arxiv.org/abs/1001.4538}{{\tt arXiv:1001.4538 [astro-ph.CO]}}.

\bibitem{Bardeen:1980kt}
J.~M. Bardeen, ``{Gauge Invariant Cosmological Perturbations}'',
\href{http://dx.doi.org/10.1103/PhysRevD.22.1882}{{\em Phys. Rev.} {\bf D22}
  (1980)  1882--1905}.

\bibitem{Mukhanov:1990me}
V.~F. Mukhanov, H.~A. Feldman, and R.~H. Brandenberger, ``{Theory of
  cosmological perturbations}'',
\href{http://dx.doi.org/10.1016/0370-1573(92)90044-Z}{{\em Phys. Rept.} {\bf
  215} (1992)  203--333}.

\bibitem{Malik:2003mv}
K.~A. Malik and D.~Wands, ``{Evolution of second-order cosmological
  perturbations}'', {\em Class. Quant. Grav.} {\bf 21} (2004)  L65--L72,
\href{http://arxiv.org/abs/astro-ph/0307055}{{\tt arXiv:astro-ph/0307055}}.

\bibitem{Malik:2005cy}
K.~A. Malik, ``{Gauge-invariant perturbations at second order: Multiple scalar
  fields on large scales}'',
  \href{http://dx.doi.org/10.1088/1475-7516/2005/11/005}{{\em JCAP} {\bf 0511}
  (2005)  005},
\href{http://arxiv.org/abs/astro-ph/0506532}{{\tt arXiv:astro-ph/0506532
  [astro-ph]}}.

\bibitem{Noh:2003yg}
H.~Noh and J.-c. Hwang, ``{Second-order perturbations of the Friedmann world
  model}'',
\href{http://arxiv.org/abs/astro-ph/0305123}{{\tt arXiv:astro-ph/0305123}}.

\bibitem{Ellis:1989jt}
G.~F.~R. Ellis and M.~Bruni, ``Covariant and gauge invariant approach to
  cosmological density flunctuations'',
\href{http://dx.doi.org/10.1103/PhysRevD.40.1804}{{\em Phys. Rev.} {\bf D40}
  (1989)  1804--1818}.

\bibitem{Rigopoulos:2005xx}
G.~I. Rigopoulos, E.~P.~S. Shellard, and B.~J.~W. van Tent, ``{Non-linear
  perturbations in multiple-field inflation}'',
  \href{http://dx.doi.org/10.1103/PhysRevD.73.083521}{{\em Phys. Rev.} {\bf
  D73} (2006)  083521},
\href{http://arxiv.org/abs/astro-ph/0504508}{{\tt arXiv:astro-ph/0504508}}.

\bibitem{Rigopoulos:2004gr}
G.~I. Rigopoulos and E.~P.~S. Shellard, ``{Non-linear inflationary
  perturbations}'', \href{http://dx.doi.org/10.1088/1475-7516/2005/10/006}{{\em
  JCAP} {\bf 0510} (2005)  006},
\href{http://arxiv.org/abs/astro-ph/0405185}{{\tt arXiv:astro-ph/0405185}}.

\bibitem{Langlois:2005qp}
D.~Langlois and F.~Vernizzi, ``{Conserved non-linear quantities in
  cosmology}'', \href{http://dx.doi.org/10.1103/PhysRevD.72.103501}{{\em Phys.
  Rev.} {\bf D72} (2005)  103501},
\href{http://arxiv.org/abs/astro-ph/0509078}{{\tt arXiv:astro-ph/0509078}}.

\bibitem{Tzavara:2010ge}
E.~Tzavara and B.~van Tent, ``{Bispectra from two-field inflation using the
  long- wavelength formalism}'',
  \href{http://dx.doi.org/10.1088/1475-7516/2011/06/026}{{\em JCAP} {\bf 1106}
  (2011)  026},
\href{http://arxiv.org/abs/1012.6027}{{\tt arXiv:1012.6027 [astro-ph.CO]}}.

\bibitem{Rigopoulos:2011eq}
G.~Rigopoulos, ``{Gauge invariance and non-Gaussianity in Inflation}'',
\href{http://arxiv.org/abs/1104.0292}{{\tt arXiv:1104.0292 [astro-ph.CO]}}.

\bibitem{Starobinsky:1986fxa}
A.~A. Starobinsky, ``{Multicomponent de Sitter (Inflationary) Stages and the
  Generation of Perturbations}'',
{\em JETP Lett.} {\bf 42} (1985)  152--155.

\bibitem{Sasaki:1995aw}
M.~Sasaki and E.~D. Stewart, ``{A General analytic formula for the spectral
  index of the density perturbations produced during inflation}'',
  \href{http://dx.doi.org/10.1143/PTP.95.71}{{\em Prog. Theor. Phys.} {\bf 95}
  (1996)  71--78},
\href{http://arxiv.org/abs/astro-ph/9507001}{{\tt arXiv:astro-ph/9507001}}.

\bibitem{Sasaki:1998ug}
M.~Sasaki and T.~Tanaka, ``{Super-horizon scale dynamics of multi-scalar
  inflation}'', \href{http://dx.doi.org/10.1143/PTP.99.763}{{\em Prog. Theor.
  Phys.} {\bf 99} (1998)  763--782},
\href{http://arxiv.org/abs/gr-qc/9801017}{{\tt arXiv:gr-qc/9801017}}.

\bibitem{Lyth:2004gb}
D.~H. Lyth, K.~A. Malik, and M.~Sasaki, ``{A general proof of the conservation
  of the curvature perturbation}'',
  \href{http://dx.doi.org/10.1088/1475-7516/2005/05/004}{{\em JCAP} {\bf 0505}
  (2005)  004},
\href{http://arxiv.org/abs/astro-ph/0411220}{{\tt arXiv:astro-ph/0411220}}.

\bibitem{Lyth:2005fi}
D.~H. Lyth and Y.~Rodriguez, ``{The inflationary prediction for primordial non-
  gaussianity}'', \href{http://dx.doi.org/10.1103/PhysRevLett.95.121302}{{\em
  Phys. Rev. Lett.} {\bf 95} (2005)  121302},
\href{http://arxiv.org/abs/astro-ph/0504045}{{\tt arXiv:astro-ph/0504045}}.

\bibitem{Maldacena:2002vr}
J.~M. Maldacena, ``{Non-Gaussian features of primordial fluctuations in single
  field inflationary models}'', {\em JHEP} {\bf 05} (2003)  013,
\href{http://arxiv.org/abs/astro-ph/0210603}{{\tt arXiv:astro-ph/0210603}}.

\bibitem{Chen:2006nt}
X.~Chen, M.~xin Huang, S.~Kachru, and G.~Shiu, ``{Observational signatures and
  non-Gaussianities of general single field inflation}'', {\em JCAP} {\bf 0701}
  (2007)  002,
\href{http://arxiv.org/abs/hep-th/0605045}{{\tt arXiv:hep-th/0605045}}.

\bibitem{Seery:2005gb}
D.~Seery and J.~E. Lidsey, ``{Primordial non-gaussianities from multiple-field
  inflation}'', \href{http://dx.doi.org/10.1088/1475-7516/2005/09/011}{{\em
  JCAP} {\bf 0509} (2005)  011},
\href{http://arxiv.org/abs/astro-ph/0506056}{{\tt arXiv:astro-ph/0506056}}.

\bibitem{Langlois:2008qf}
D.~Langlois, S.~Renaux-Petel, D.~A. Steer, and T.~Tanaka, ``{Primordial
  perturbations and non-Gaussianities in DBI and general multi-field
  inflation}'', \href{http://dx.doi.org/10.1103/PhysRevD.78.063523}{{\em Phys.
  Rev.} {\bf D78} (2008)  063523},
\href{http://arxiv.org/abs/0806.0336}{{\tt arXiv:0806.0336 [hep-th]}}.

\bibitem{Arroja:2008yy}
F.~Arroja, S.~Mizuno, and K.~Koyama, ``{Non-gaussianity from the bispectrum in
  general multiple field inflation}'',
  \href{http://dx.doi.org/10.1088/1475-7516/2008/08/015}{{\em JCAP} {\bf 0808}
  (2008)  015},
\href{http://arxiv.org/abs/0806.0619}{{\tt arXiv:0806.0619 [astro-ph]}}.

\bibitem{Gao:2008dt}
X.~Gao, ``{Primordial Non-Gaussianities of General Multiple Field Inflation}'',
  \href{http://dx.doi.org/10.1088/1475-7516/2008/06/029}{{\em JCAP} {\bf 0806}
  (2008)  029},
\href{http://arxiv.org/abs/0804.1055}{{\tt arXiv:0804.1055 [astro-ph]}}.

\bibitem{Weinberg:2005vy}
S.~Weinberg, ``{Quantum contributions to cosmological correlations}'',
  \href{http://dx.doi.org/10.1103/PhysRevD.72.043514}{{\em Phys. Rev.} {\bf
  D72} (2005)  043514},
\href{http://arxiv.org/abs/hep-th/0506236}{{\tt arXiv:hep-th/0506236}}.

\bibitem{Misner:1974qy}
C.~W. Misner, K.~S. Thorne, and J.~A. Wheeler, ``{Gravitation}'',. San
  Francisco 1973, 1279p.

\bibitem{GrootNibbelink:2001qt}
S.~{Groot Nibbelink} and B.~J.~W. van Tent, ``{Scalar perturbations during
  multiple field slow-roll inflation}'',
  \href{http://dx.doi.org/10.1088/0264-9381/19/4/302}{{\em Class. Quant. Grav.}
  {\bf 19} (2002)  613--640},
\href{http://arxiv.org/abs/hep-ph/0107272}{{\tt arXiv:hep-ph/0107272}}.

\bibitem{Lyth:2005du}
D.~H. Lyth and Y.~Rodriguez, ``{Non-Gaussianity from the second-order
  cosmological perturbation}'',
  \href{http://dx.doi.org/10.1103/PhysRevD.71.123508}{{\em Phys.Rev.} {\bf D71}
  (2005)  123508},
\href{http://arxiv.org/abs/astro-ph/0502578}{{\tt arXiv:astro-ph/0502578
  [astro-ph]}}.

\bibitem{Bruni:1996im}
M.~Bruni, S.~Matarrese, S.~Mollerach, and S.~Sonego, ``{Perturbations of
  space-time: Gauge transformations and gauge invariance at second order and
  beyond}'', \href{http://dx.doi.org/10.1088/0264-9381/14/9/014}{{\em
  Class.Quant.Grav.} {\bf 14} (1997)  2585--2606},
  \href{http://arxiv.org/abs/gr-qc/9609040}{{\tt arXiv:gr-qc/9609040 [gr-qc]}}.

\bibitem{Salopek:1990re}
D.~S. Salopek and J.~R. Bond, ``{Stochastic inflation and nonlinear gravity}'',
\href{http://dx.doi.org/10.1103/PhysRevD.43.1005}{{\em Phys. Rev.} {\bf D43}
  (1991)  1005--1031}.

\bibitem{Rigopoulos:2005us}
G.~I. Rigopoulos, E.~P.~S. Shellard, and B.~J.~W. van Tent, ``{Quantitative
  bispectra from multifield inflation}'',
  \href{http://dx.doi.org/10.1103/PhysRevD.76.083512}{{\em Phys. Rev.} {\bf
  D76} (2007)  083512},
\href{http://arxiv.org/abs/astro-ph/0511041}{{\tt arXiv:astro-ph/0511041}}.

\bibitem{Seery:2005wm}
D.~Seery and J.~E. Lidsey, ``{Primordial non-Gaussianities in single field
  inflation}'', \href{http://dx.doi.org/10.1088/1475-7516/2005/06/003}{{\em
  JCAP} {\bf 0506} (2005)  003},
\href{http://arxiv.org/abs/astro-ph/0503692}{{\tt arXiv:astro-ph/0503692
  [astro-ph]}}.

\end{thebibliography}\endgroup

\bibliographystyle{utphys.bst}

\end{document}